\documentclass[12pt]{article}
\usepackage[a4paper,margin=1in]{geometry}
\usepackage{amsmath,amssymb,bm}
\usepackage{siunitx}
\usepackage{booktabs}
\usepackage{graphicx}
\usepackage{enumitem}
\usepackage{amsmath}
\usepackage{caption,subcaption}
\usepackage{float}
\usepackage{tikz}
\usetikzlibrary{decorations.pathmorphing}
\usepackage{hyperref}
\usepackage{setspace}
\title {Theory of Handedness Selection in Helices of Chiral Polymers and Biopolymers}

\author{Biman Bagchi*\\
Solid State and Structural Chemistry Unit\\
Indian Institute of Science, Bengaluru 560012, India.\\
\texttt{profbiman@gmail.com}}

\date{}
\begin{document}

\maketitle

\begin{abstract}
Helices are among the most common ordered structures in biological and
synthetic polymers, but their formation involves more than local conformational
preference. A finite helix must nucleate, grow, resist breaking, and maintain a
selected handedness against thermal fluctuations. We develop a three-state
Ising-like transfer-matrix theory in which each segment is coil-like,
right-handed helical, or left-handed helical. This formulation separates
ordinary helix--coil cooperativity from the persistence of handedness. The
right--left symmetric problem decomposes into symmetric and antisymmetric
sectors, giving two characteristic lengths: the helical correlation length
$\xi_H$ and the chiral persistence length $\xi_\chi$. In the strongly helical
rare-wall limit, $\xi_\chi\simeq (1/2)\exp[\beta(K+J)]$. A local chiral bias,
motivated by the Ramachandran landscape, is then amplified over finite helical
domains.

\end{abstract}

\section{Introduction}

An important problem in molecular biology is to understand how local interactions between neighboring constituents give rise to extended ordered domains. In polymers, this problem is considerably richer because the ordered helical state competes continuously with the entropically favored coil state. Moreover, opposite chiral domains cannot generally interconvert directly but must pass through local unwinding into coil configurations before reforming with the opposite handedness. Understanding how weak molecular biases are cooperatively amplified into finite chiral domains requires a theory that combines cooperativity, entropy, and kinetics within a unified statistical-mechanical framework. Despite advances in helix-coil transition theories, 
such a framework is missing.

The pioneering work of Pauling and Corey established the $\alpha$-helix as a fundamental structural motif in proteins and opened the way to a statistical-mechanical view of helical order in polypeptide chains.\cite{PaulingCoreyBranson1951} Subsequent biochemical and polymer-physical descriptions made clear that helices are not rigid crystalline objects. As emphasized in standard biochemical treatments, including the classic text by Lehninger,\cite{Lehninger} and in later structural discussions,\cite{BrandenTooze} a helix is a soft, fluctuating, and finite object whose stability reflects a balance between local energetic stabilization and conformational entropy. It can breathe, fray at its ends, break, reform, and migrate along the chain. Thus the problem of a finite helix is not only the problem of forming a locally ordered structure, but also the problem of understanding how that order is cooperatively stabilized and how it is lost.

For proteins, the local stereochemical preference for the right-handed $\alpha$-helical basin is well understood. The Ramachandran plot maps the allowed backbone torsional angles $(\phi,\psi)$ and shows how the stereochemistry of L-amino acids, together with steric constraints, hydrogen-bond compatibility, and side-chain packing, favors the right-handed $\alpha$-helical region over its left-handed counterpart.\cite{Lehninger,Ramachandran1963} Microscopic theories of chain conformation and helix--coil transitions, including the Gibbs-DiMarzio cooperative hydrogen bond breaking theory,  the Zimm--Bragg transfer matrix theory, the Lifson--Roig theory, and Flory's rotational-isomeric-state approach, have provided powerful descriptions of local conformational statistics, nucleation, propagation, and coil entropy.\cite{GibbsDiMarzio1958, ZimmBragg1958,ZimmBragg1959, LifsonRoig1961,FloryRIS,PolandScheraga1970} Therefore, the question addressed here is not whether a right-handed $\alpha$-helix is locally feasible. Rather, the question is how local helical order and local chiral preference are amplified, protected, and eventually destroyed in a finite fluctuating domain.


This distinction is important because a finite helical segment is not stabilized by a single interaction alone. Its formation and persistence reflect a cooperative balance among local ordering, hydrogen bonding or sticker interactions, bending and torsional costs, packing forces, side-chain contacts, solvent effects, and the entropy of the competing coil state. A segment of length $n$ is therefore a collective object. Its stability cannot be characterized only by the free energy of one residue or one bond; it also depends on the cost of creating helix--coil boundaries, the free energy gained by same-handed propagation, and the defects that break the continuity of the helical state.

Chirality adds a further level of cooperativity. A helical segment is not only ordered; it also has a handedness. In proteins the local bias toward right-handed $\alpha$-helical geometry is already present at the level of the allowed torsional landscape, but the persistence of that handedness over many residues is a separate statistical-mechanical question. A reversal of handedness inside an otherwise helical domain is not an ordinary continuation of the same helix. It disrupts torsional registry, hydrogen-bonding patterns, packing contacts, and local solvation, and therefore acts as a domain-wall-like defect. Thus a theory of finite helices should describe not only helix formation, but also helix breaking, handedness reversal, and the migration of helical domains along the chain.

Helical organization is ubiquitous in biological and synthetic systems. In biological macromolecules, helices occur in proteins, RNA, DNA, amylose, and numerous filamentous assemblies.\cite{ImbertyPerez,TinocoBustamante} Double-helical DNA represents a stable and highly organized example of molecular chirality, while RNA and protein helices often occur as finite, fluctuating, environment-dependent motifs. Helical structures also arise in synthetic polymers such as polyisocyanates, substituted polyacetylenes, polypeptides, foldamers, and peptoid systems.\cite{GreenEtAl,DotyYang,NakanoFujiki,GellmanFoldamers} In many of these systems, the relevant physical object is not an infinitely long perfect helix, but a finite helical domain whose length, handedness, lifetime, and mobility are determined by cooperative interactions.

This viewpoint is closely related to the broader problem of polymer collapse and morphology selection. A semiflexible polymer in a poor solvent may collapse into several competing morphologies, including disordered globules, toroids, rods, and other compact states. In recent work, the free-energy landscape of such semiflexible polymers was analyzed in terms of the competition among attraction, bending stiffness, orientational ordering, and surface cost, \cite{BagchiMorphology2026}, further discussed below. 

Several recent and classical perspectives help clarify why a statistical-mechanical treatment of handedness in finite helices is needed. Efimov has emphasized that protein structures exhibit handedness at many levels of organization, including $\alpha$-helices, $\beta$-strands, hairpins, $\beta\alpha\beta$ units, superhelices, and more complex structural motifs; he also distinguishes ordinary molecular chirality from the handedness, or ``pseudochirality,'' of polypeptide conformations, since left- and right-handed conformational forms can interconvert without conversion of L-amino acids into D-amino acids.\cite{Efimov2018} This viewpoint supports the present distinction between local stereochemical bias and the cooperative persistence of handedness in a finite helical domain. A complementary lesson comes from the recent work of Roy, Appadurai, and Srivastava on kinked-$\beta$ sheets, where motifs that deviate from standard $\beta$-strand regions nevertheless remain distributed within the broader allowed regions of Ramachandran space and require additional structural descriptors, such as bend and rotation angles, for unambiguous classification.\cite{RoyAppaduraiSrivastava2025} This illustrates a general point relevant here: the Ramachandran plot identifies locally allowed torsional regions, but additional collective or mesoscopic variables are often needed to describe structural defects, reversibility, and dynamical persistence. Finally, the concept of degeneracy discussed by Edelman and Gally is useful in interpreting the effective parameters of the present model.\cite{EdelmanGally2001} In biological systems, structurally different elements may produce similar functional outcomes, a property distinct from simple redundancy. 

Three recent studies form the immediate background to the present work. In
the first, we examined the morphology selection of semiflexible polymers
under collapse and analyzed the competition among globular, toroidal,
rod-like, and related compact structures.\cite{BagchiMorphology2026} That
study reinforced an important distinction: although ordinary bending
elasticity, surface free energy, and isotropic attraction can account for
several compact polymer morphologies, they do not generically select a
helical structure.

In a subsequent study, we addressed explicitly the microscopic emergence of
helices in polymer condensates.\cite{BagchiEmergingHelices2026} Two minimal
routes to helix stabilization were identified. The first is geometric:
packing constraints in a thick or semiflexible chain can select a structure
with finite curvature, torsion, radius, and pitch. The second is
interaction-specific: hydrogen bonding or periodically spaced sticky groups
can favor a commensurate registry of contacts along the chain. This work
showed why helices require additional geometric or directional interactions
beyond those sufficient to produce globules, toroids, and rods. However,
right- and left-handed helices remained degenerate, so the theory did not
provide a mechanism for reproducible handedness selection.

More recently, a
statistical-mechanical and dynamical theory of the cooperative nucleation,
growth, fluctuation, and lifetime of finite helical segments was developed.\cite{BagchiLivingHelices2026}
Helix formation was described as a multistep process involving a constrained
pre-nucleus followed by cooperative stabilization and growth. The resulting
helices are finite, mobile, and dynamically fluctuating rather than permanent
ordered structures. That theory treated the emergence and dynamics of
helical segments in greater detail, but remained symmetric under interchange
of right- and left-handed helices and contained no intrinsic chiral field.
The present work completes this progression by treating helical occupancy
and handedness as distinct collective variables and by examining chiral-wall
formation, handedness correlations, and the cooperative amplification of a
weak stereochemical bias.

Classical theories of helix--coil transitions, especially those of Zimm and
Bragg, Gibbs and DiMarzio, Lifson and Roig, and Poland and Scheraga, provide a
powerful foundation for describing cooperative helix formation
\cite{GibbsDiMarzio1958, ZimmBragg1958,LifsonRoig1961, PolandScheraga1970}.
These theories introduced the central ideas of propagation, nucleation, and
coil entropy.  However, in their simplest forms they do not explicitly separate
the helical occupancy of a segment from its handedness.  Consequently, they do
not directly yield separate analytical expressions for a helical correlation
length and a chiral persistence length.

It is useful to state the meaning of the principal parameters and length
scales of the theory at the outset. The quantity $\sigma$ denotes the usual
helix--coil nucleation, or cooperativity, parameter of the Zimm--Bragg and
related Lifson--Roig descriptions.\cite{ZimmBragg1958, ZimmBragg1959, LifsonRoig1961}
Physically, $\sigma$ measures the free-energy penalty for creating the two 
helix--coil boundaries that enclose a finite helical domain within the coil. 
A small $\sigma$ therefore suppresses isolated helical residues, while allowing 
favorable growth once a nucleus has formed. 

The helical correlation length $\xi_H$ measures the typical distance over which
helical order persists along the chain. Near the cooperative helix--coil
crossover, nucleation is strongly suppressed when $\sigma \ll 1$, so that once
a helical segment forms it tends to propagate over many residues before
terminating. Because the chain is effectively one-dimensional and the dominant
defects are the two helix--coil boundaries, the Zimm--Bragg transfer-matrix result gives \cite{ZimmBragg1959}
\begin{equation}
\xi_H \sim \sigma^{-1/2}.
\end{equation}

The persistence of handedness is controlled by a different free-energy scale.
We denote by $K>0$ the magnitude of the free-energy stabilization gained when
two neighboring helical segments propagate with the same handedness. The
corresponding contribution to the Hamiltonian therefore lowers the free
energy of a same-handed helical contact. We denote by $J>0$ the additional
free-energy penalty associated with placing two neighboring helical segments
in opposite handedness. Replacing a same-handed contact by a wrong-handed
contact thus loses the propagation stabilization $K$ and incurs the additional
penalty $J$. The resulting chiral-domain-wall free energy is therefore $\Delta F_{\rm wall}=K+J$.

The chiral correlation length $\xi_\chi$ measures the distance over which a
selected handedness remains correlated. In the strongly helical, rare-wall
limit, the present theory gives
\begin{equation}
\xi_\chi\simeq \frac{1}{2}\exp[\beta(K+J)],
\qquad
\beta=(k_{\rm B}T)^{-1}.
\end{equation}
Thus, $\xi_H$ characterizes the persistence of helical occupancy, whereas
$\xi_\chi$ characterizes the persistence of handedness within the helical
state. These two length scales arise from different physical mechanisms and
need not be comparable.


For protein-like helices, a chiral-domain-wall free energy of only a few
$k_{\rm B}T$ is already sufficient to produce handedness correlations over
tens of residues. In the strongly helical rare-wall regime, inversion of the
preceding asymptotic result gives the estimate
\[
\frac{\Delta F_{\rm wall}}{k_{\rm B}T}
\simeq \ln(2\xi_\chi).
\]
A chiral correlation length of 10--100 segments therefore corresponds to a
wall free energy of approximately $3$--$5\,k_{\rm B}T$. Such values are
physically plausible when torsional frustration, hydrogen-bond registry,
side-chain packing, hydrophobic interactions, and solvent-mediated forces
contribute collectively to the effective wall free energy.

The length scales addressed by the theory are experimentally relevant.
Structural surveys show that ordinary protein $\alpha$-helices are small, found to be frequently
of order ten residues in length,\cite{KabschSander1983,Creighton1993} while
helix--coil kinetic studies of alanine-based peptides have examined well-defined
helical segments containing 19--39 residues.\cite{HuangGetahunZhuKallenbach2004}
Still longer single-$\alpha$-helical domains are also known. For example, NMR
studies of the medial tail domain of myosin VI found an ordered helix extending
from Glu-6 to Lys-63 and a mechanical persistence length of
$224\pm10$~\AA{} at $20\,^\circ$C.\cite{BarnesMyosinVI2019} This mechanical
persistence length is not identical to the chiral correlation length
$\xi_\chi$, but it demonstrates that long, structurally coherent
single-helical domains can occur in proteins. Helical and chiral domains
spanning approximately 10--100 residues are therefore physically meaningful
target scales for the present theory.

Sequence heterogeneity is, of course, central to helix formation in proteins:
different residues contribute differently to hydrogen bonding, hydrophobic
packing, side-chain entropy, salt bridges, and solvent exposure.  The present
model does not attempt a residue-specific theory, but incorporates such effects
through effective parameters that may later be estimated from experiment,
simulation, or sequence-dependent scales.


\begin{figure}[t]
\centering
\includegraphics[width=0.95\linewidth]{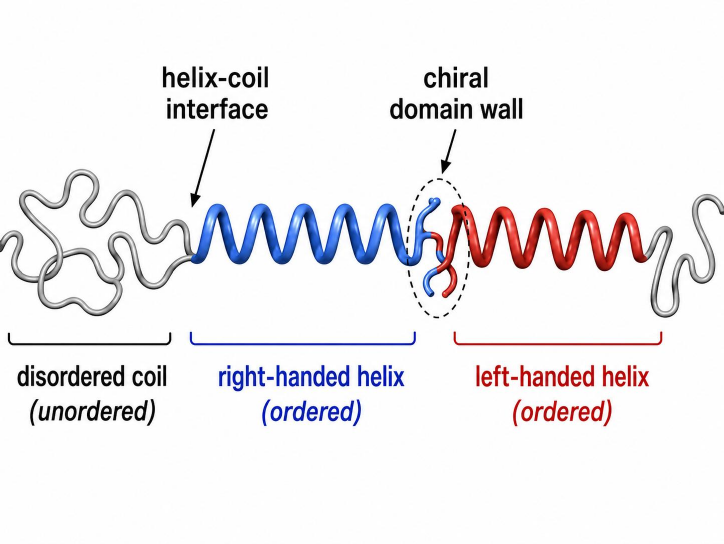}
\caption{Schematic illustration of the three local environments considered in the present theory.  A disordered coil segment (gray) is separated from an ordered helical segment by a helix--coil interface.  The two possible helical senses, shown here as right-handed (blue) and left-handed (red), are separated by a chiral domain wall.  The figure emphasizes the central physical distinction of the model: a helix--coil boundary is different from a handedness-reversal defect within an otherwise helical region.  In the theory, \(K\) denotes the free-energy stabilization of same-handed helical propagation, while \(J\) denotes the additional penalty associated with a wrong-handed contact; thus a chiral domain wall costs the lost propagation free energy plus the explicit wall penalty.}
\label{fig:helix_coil_domain_wall}
\label{fig:placeholder}
\end{figure}

An important experimental development beyond the classical helix--coil theories is the demonstration that polymers can exhibit pronounced chiral amplification, whereby a small molecular or conformational bias is cooperatively amplified into a nearly single-handed macroscopic helix. Green et al. showed that a slight excess of one chiral component in suitably designed polymers can produce an overwhelmingly preferred helical sense through cooperative interactions.\cite{GreenEtAl}

\textit{This phenomenon highlights the remarkable sensitivity of helical systems to weak chiral biases and provides experimental evidence that handedness selection need not arise from a large intrinsic asymmetry.} The present theory explores an analogous question from a statistical-mechanical perspective by introducing an intrinsic local chiral bias $\Delta_\chi$ (which is torsional in original), whose cooperative amplification along the chain gives rise to stable right- or left-handed helical domains and their transitions.

Although there have been computer simulation studies of the formation and
melting of helices \cite{Jana1,Hummer2000}, there have been fewer theoretical
studies of chirality and handedness, especially starting from the Ramachandran
plot. In particular, Hummer, Garc\'ia, and Garde showed that helix formation in
short peptides can be viewed as conformational diffusion within the coil
ensemble, followed by rapid ordering into the helical state; their analysis also
emphasized local nucleation, end fraying, and hydrogen-bond rearrangement in
helix formation and melting, rather than the separate problem of handedness
persistence.\cite{Hummer2000}

In this work we develop a minimal three-state transfer-matrix theory that
retains helical occupancy and handedness as distinct variables. Each polymer
segment is assigned a local state $s_i=0,+1,-1$, where $s_i=0$ denotes coil
and $s_i=\pm1$ denote right- and left-handed helical states. The helical
occupancy is represented by $n_i=s_i^2$, whereas $s_i$ itself carries the
local handedness. The large conformational entropy of the coil state is
included through an effective degeneracy factor $g$. Same-handed helical
propagation is stabilized by $K$, while a wrong-handed neighboring contact
carries the additional penalty $J$. The model therefore distinguishes an
ordinary helix--coil boundary from a handedness-reversal wall within an
otherwise helical domain and produces separate helical and chiral correlation
lengths, $\xi_H$ and $\xi_\chi$.

 In the absence of an
intrinsic chiral bias, it possesses an exact right--left exchange symmetry.
Under the transformation $s_i\rightarrow -s_i$, the right- and left-handed
helical states are interchanged while the coil state remains unchanged. This
defines a discrete $\mathbb{Z}_2$ symmetry analogous to the spin-reversal
symmetry of the Ising model. Although each segment has three local states
rather than two, the transfer-matrix spectrum separates into symmetric and
antisymmetric sectors: fluctuations of helical occupancy are associated
primarily with the symmetric sector, whereas handedness correlations are
governed by the antisymmetric sector. The resulting Ising-like behavior
therefore arises from the symmetry of the handedness variable, rather than
from the number of local conformational states.\cite{Yeomans,Cardy}

The analogy with a ferromagnetic Ising system is particularly useful
physically. The combined wall free energy $K+J$ suppresses reversals of
handedness in the same way that a ferromagnetic interaction suppresses spin
domain walls. The intrinsic chiral bias $\Delta_\chi$, introduced later,
plays the role of an external field. When $\Delta_\chi=0$, the two handedness
states are degenerate and the symmetric and antisymmetric transfer-matrix
sectors remain distinct. When $\Delta_\chi\neq0$, the right--left exchange
symmetry is explicitly broken, the two sectors are coupled, and one helical
sense is thermodynamically selected. The cooperative interactions do not
create the sign of the preference; rather, they amplify and preserve the
small local bias over a finite helical domain.

More generally, a local perturbation propagated over a separation $j$
contains contributions proportional to
$(\lambda_\alpha/\lambda_+)^j$, where $\lambda_+$ is the dominant
transfer-matrix eigenvalue and $\lambda_\alpha$ is a subleading eigenvalue.
The corresponding correlation lengths are therefore determined by ratios of
transfer-matrix eigenvalues. At sufficiently large separation, the slowest
symmetry-allowed subleading mode dominates, although the approach to this
asymptotic regime depends on the separation of the eigenvalues and on the
overlap of the observable with the corresponding mode.\cite{Yeomans,Baxter}

The rest of the paper is organized as follows. In Sec.~2 we discuss the
thermodynamic basis of the three-state model and the role of coil entropy.
In Sec.~3 we identify the physical contributions to the effective helix,
same-handed propagation, and chiral-domain-wall free energies. In Sec.~4 we
introduce the three-state Hamiltonian and the projectors that distinguish
helical occupancy, same-handed propagation, and wrong-handed contacts.
In Sec.~5 we construct and diagonalize the transfer matrix and discuss the
right--left exchange symmetry of the unbiased model. In Sec.~6 we derive the
helical and chiral correlation lengths and examine their dependence on the
model parameters. Section~7 applies the theory to representative protein
helices and compares the predicted length scales with experimental estimates.
Section~8 discusses chiral-wall dynamics, domain lifetimes, and the migration
of finite helical domains. In Sec.~9 we introduce the intrinsic chiral bias,
show how it breaks the right--left symmetry, and quantify the cooperative
selection of handedness in proteins and synthetic helical polymers. We
summarize the principal results and their physical implications in Sec.~10,
while additional parameter estimates and limiting results are collected in
the Appendix.


\section{Physical Basis of the Three-State Model}

The formation of a helical segment in a flexible polymer is governed by
the competition between energetic stabilization and conformational
entropy. A coil-like segment represents a large ensemble of microscopic
backbone, side-chain, and solvent-coupled conformations. We denote the
effective number of such conformations per coarse-grained segment by
$\Omega_c$ and define the corresponding coil entropy per segment as $s_c = k_B \ln \Omega_c $.

The local free energy of a coil-like segment is therefore written as $f_c(T)=e_c-Ts_c$,
where $e_c$ denotes its effective energetic contribution. Similarly, the
local free energy of a helical segment is
$f_h(T)=e_h-Ts_h$,
where $s_h$ is the conformational entropy of the helical state.

The local helix--coil free-energy difference is then defined as
\begin{equation}
\Delta f_{hc}(T)
\equiv f_h(T)-f_c(T)
=
\Delta e_{hc}-T\Delta s_{hc},
\end{equation}
with
\begin{equation}
\Delta e_{hc}=e_h-e_c,
\qquad
\Delta s_{hc}=s_h-s_c.
\end{equation}
Since the coil state generally possesses the larger conformational
entropy, $s_c>s_h$, one normally has $\Delta s_{hc}<0$. Helix formation
is locally favorable when $\Delta f_{hc}(T)<0 $.

Consider a long polymer chain containing $N$ coarse-grained segments,
with $N\gg 1$, and suppose that a helical domain of length $n$ is
nucleated within the coil background. Let $\gamma_{hc}$ denote the
free-energy cost of one helix--coil boundary. If multiple helical domains
are sufficiently well separated that their mutual interactions may be
neglected, the free-energy change associated with forming a single
helical domain is
\begin{equation}
\Delta F(n)
=
2\gamma_{hc}
+
n\,\Delta f_{hc}(T).
\end{equation}
The first term is the free-energy cost of the two helix--coil boundaries,
whereas the second term is the bulk free-energy change associated with
converting $n$ coil-like segments into helical segments.
This equation is the one-dimensional analogue of the familiar
surface--bulk competition in classical nucleation theory. When
$\Delta f_{hc}(T)<0$, extension of a helical domain is thermodynamically
favored after the boundary cost has been overcome. The corresponding
crossover size is
\begin{equation}
n^{*}
=
\frac{2\gamma_{hc}}
{-\Delta f_{hc}(T)},
\qquad
\Delta f_{hc}(T)<0.
\end{equation}
Near the helix--coil crossover, $\Delta f_{hc}(T)$ is small in magnitude,
so that even a moderate helix--coil boundary free energy produces
strongly cooperative helix formation.

This thermodynamic picture provides the physical basis of the classical
helix--coil theories of Zimm and Bragg, Gibbs and DiMarzio, Lifson and
Roig, and Poland and Scheraga. In the present notation, the
Zimm--Bragg-like cooperativity parameter may be related to the cost of
creating the two ends of a helical domain through
\begin{equation}
\sigma
=
\exp\left(-2\beta\gamma_{hc}\right),
\end{equation}
where \(\gamma_{hc}>0\) is the free-energy cost of a single helix--coil
boundary, so that \(2\gamma_{hc}\) is the total boundary cost for a finite
helical domain embedded in the coil state.
The helix--coil boundary free energy determines the nucleation
penalty and the degree of cooperativity. The present theory retains this
physical interpretation while explicitly separating the coil entropy
$s_c$, the coil free energy $f_c$, and the local helix--coil free-energy
difference $\Delta f_{hc}(T)$.

The starting point of the present theory is the Ramachandran conformational
landscape. Instead of treating the backbone dihedral angles as continuous
variables, we partition the $(\phi,\psi)$ space into a small number of
physically relevant basins. 
The large ensemble of non-helical conformations is represented by one
coarse-grained coil state whose statistical weight is determined by its
free energy $f_c(T)=e_c-Ts_c$, whereas the right- and left-handed helical
basins are represented by two discrete ordered states.
This coarse graining preserves the essential
competition between the entropy of the coil ensemble and the energetic
stabilization of helical conformations while permitting a simple analytical
description based on transfer matrices.


\begin{figure}[htbp]
    \centering
    \includegraphics[
        width=0.6\linewidth,
    ]{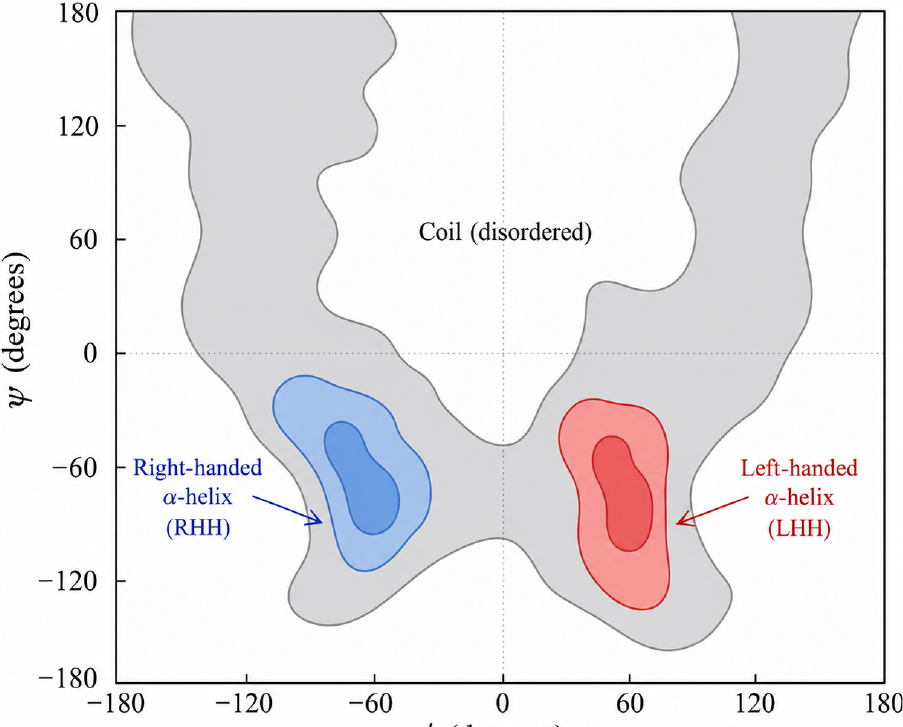}
    \caption{Ramachandran plot showing the sterically allowed regions of the backbone dihedral angles $\phi$ and $\psi$. The most favored regions correspond to common secondary structures, including $\alpha$-helices and $\beta$-sheets, while additional allowed regions represent less frequently adopted conformations.}

    \label{fig:Ramachandran}
\end{figure} 

A consequence of this coarse graining is that the helical manifold
contains two symmetry-related states corresponding to opposite handedness.
Their explicit inclusion leads naturally to a minimal three-state description,
which distinguishes the formation of helical domains from the persistence of their handedness. As will be shown below, these two collective phenomena are governed by
different thermodynamic mechanisms. Helix formation is controlled by
the local free-energy difference $\Delta f_{hc}(T)$ together with the
helix--coil boundary free energy $\gamma_{hc}$, whereas the persistence
of handedness is controlled by domain-wall fluctuations within the
helical manifold.


\section{Competing Forces in Helix Formation and Chiral Domain Selection}

The entropic advantage of the coil state originates from the large
region of configurational space accessible to a coil-like segment. We
denote the effective number of microscopic coil configurations
represented by one coarse-grained coil state by $g$. The quantity $g$
is therefore a dimensionless configurational multiplicity, not an
entropy. The corresponding coil entropy per coarse-grained segment is
defined as $ s_c = k_B \ln g $. The entropic contribution to the coil free energy is consequently
$-Ts_c = -k_BT\ln g$. If the entropy of a helical segment is denoted by $s_h$, the entropy
lost upon converting a coil-like segment into a helical segment is
$\Delta s_{\mathrm{loss}} = s_c-s_h > 0$ .
The associated free-energy penalty is $T\Delta s_{\mathrm{loss}}
= T(s_c-s_h)$.

In a simplified description in which the residual entropy of the
helical state is neglected, $s_h\simeq 0$, this reduces to
\begin{equation}
T\Delta s_{\mathrm{loss}}
\simeq Ts_c
=
k_BT\ln g .
\end{equation}

A helical conformation also carries elastic costs.  For a space curve with
curvature $\kappa$ and torsion $\tau$, a simple local elastic free-energy
density may be written as

\begin{equation}
f_{\rm el}
=
\frac{\kappa_b}{2}(\kappa-\kappa_0)^2
+
\frac{\kappa_t}{2}(\tau-\tau_0)^2 .
\label{eq:elastic_cost_general}
\end{equation}

Here $\kappa_b$ and $\kappa_t$ are bending and torsional stiffnesses, while
$\kappa_0$ and $\tau_0$ denote the locally preferred curvature and torsion.  For
a helix of radius $R$ and pitch $P$, defining
$a=\frac{P}{2\pi}$, one has
\begin{equation}
\kappa=\frac{R}{R^2+a^2},
\label{eq:helix_curvature}
\end{equation}

and,

\begin{equation}
\tau=\frac{a}{R^2+a^2}.
\label{eq:helix_torsion}
\end{equation}

Right- and left-handed helices correspond to opposite signs of the torsion.  In
an achiral environment the states with $\tau=+\tau_0$ and $\tau=-\tau_0$ are
degenerate.  Thus the elastic theory does not by itself choose a global
handedness.  However, a domain wall between opposite handedness still requires a local
change in torsional registry and generally a distortion of curvature.  This
distortion contributes to the free-energy penalty for a handedness reversal.

Right- and left-handed helices correspond to opposite signs of the torsion.
In an achiral environment the two states with $\tau=+\tau_0$ and
$\tau=-\tau_0$ are degenerate.  Thus the elastic energy alone does not select a
global handedness.  This degeneracy is important: the chain may locally choose
either helical sense.  However, a domain wall between opposite handedness is not
a simple continuation of the same helical structure.  It requires a local change
of torsional registry and usually also a distortion of curvature.  Such a
distortion can break or weaken hydrogen-bond registry, disturb side-chain
packing, and reorganize the local solvent environment.  Therefore the elastic
description naturally contributes not only to the cost of forming a helix, but
also to the penalty for reversing handedness inside a helical domain.

A local helical unit stabilized by
hydrogen bonding or by a sticker interaction may be assigned a free energy contribution
\begin{equation}
\Delta f_{\rm HB}
=
-\epsilon_{\rm HB}
+
\Delta f_{\rm geom}
-
T\Delta s_{\rm coil}.
\label{eq:df_HB}
\end{equation}
Here $\epsilon_{\rm HB}>0$ is the stabilizing hydrogen-bond or sticker energy.
The term $\Delta f_{\rm geom}$ denotes the geometric cost of realizing the
helical arrangement, including bending and torsional distortions.  The final
term is the entropy loss associated with converting a coil-like segment into an
ordered helical segment.  This expression is schematic, but it is useful because
it displays the basic competition: hydrogen bonding favors helicity, whereas
geometry and coil entropy oppose it.


Packing interactions provide a second important route to helical stabilization.\cite{BagchiEmergingHelices2026}
A semiflexible or thick chain may pack more efficiently in a helical arrangement
than in a disordered local structure.  In proteins or substituted polymers, side
chains or pendant groups may also form favorable hydrophobic, aromatic,
electrostatic, or steric contacts in a regular helical geometry.  A schematic
packing contribution may now be written as
\begin{equation}
\Delta f_{\rm pack}
=
f_{\rm bend}
+
f_{\rm tors}
+
f_{\rm surf}
-
z_p\epsilon_{\rm pack}
-
T\Delta s_{\rm coil}.
\label{eq:df_pack}
\end{equation}
Here $z_p$ is the effective number of favorable packing contacts per segment and
$\epsilon_{\rm pack}$ is the average packing energy per contact.  The term
$f_{\rm surf}$ represents local solvation or surface costs.  As in the
hydrogen-bonding case, packing stabilizes a regular helical structure but also
penalizes handedness reversal, because a reversal interrupts the repeated
packing pattern.

These considerations motivate the use of effective parameters.  The local
free-energy cost of forming a helical segment may be written as
\begin{equation}
\epsilon_{\rm eff}
=
f_{\rm bend}
+
f_{\rm tors}
+
f_{\rm surf}
-
\epsilon_{\rm HB}^{\rm eff}
-
z_p\epsilon_{\rm pack}
-
\epsilon_{\rm side}
-
T\Delta s_{\rm coil}.
\label{eq:epsilon_eff}
\end{equation}
Here $\epsilon_{\rm HB}^{\rm eff}$ is the effective hydrogen-bond or sticker
stabilization, and $\epsilon_{\rm side}$ collects side-chain, aromatic,
electrostatic, hydrophobic, and solvent-mediated stabilization.  The sign
convention is such that a smaller $\epsilon_{\rm eff}$ favors helix formation.
Thus $\epsilon_{\rm eff}$ is not a microscopic energy.  It is a
coarse-grained free energy that already contains the local energetic and
entropic contributions associated with making a helical unit.

The same-handed propagation parameter \(K\) is an effective free-energy
stabilization that may contain contributions from hydrogen-bond or sticker
registry, packing, torsional continuity, side-chain interactions, and
solvent-mediated effects:
\begin{equation}
K =
K_{\rm HB}+K_{\rm pack}+K_{\rm tors}+K_{\rm side}+K_{\rm solv}.
\end{equation}

The term $K_{\rm HB}$ represents cooperative hydrogen-bond or sticker registry.
The term $K_{\rm pack}$ represents the repeated favorable packing contacts
obtained by extending the helix with the same handedness.  The term
$K_{\rm tors}$ represents the advantage of maintaining a uniform torsional
state.  The remaining terms collect side-chain and solvent-mediated
contributions.  Thus $K_{\rm eff}$ measures the free-energy gain of adding a
helical unit without changing handedness.

Similarly, the additional free-energy penalty \(J\) for a wrong-handed
neighboring contact may be written as
\begin{equation}
J
=
J_{\rm HB}
+
J_{\rm pack}
+
J_{\rm tors}
+
J_{\rm side}
+
J_{\rm solv}.
\label{eq:J}
\end{equation}

This parameter measures the additional cost of placing neighboring helical units in opposite handedness.  
Such a contact may break hydrogen-bond registry,
frustrate packing, introduce a torsional defect, disturb side-chain contacts, or
reorganize solvent.  These contributions need not be large individually.  Since
they add at the level of free energy, their combined effect can nevertheless
strongly suppress handedness reversals.

The full free-energy cost of replacing a same-handed helical contact
by a wrong-handed one is therefore
\begin{equation}
\Delta F_{\rm wall}=K+J.
\label{eq:Fwall}
\end{equation}
This expression has a simple interpretation. A wrong-handed contact
loses the same-handed propagation stabilization \(K\) and also pays
the additional mismatch penalty \(J\).
 This wall free energy is the physical basis of local homochirality in the
present model.

The probability of a handedness-reversal wall is correspondingly small:
\begin{equation}
p_{\rm wall}
\sim
\exp[-\beta \Delta F_{\rm wall}].
\label{eq:pwall_eff}
\end{equation}
Thus a helical nucleus may choose either handedness in an achiral environment,
but after one handedness has been selected, reversals are rare if
$\Delta F_{\rm wall}$ is several $k_BT$.  This is the mechanism by
which finite homochiral domains emerge without imposing an external chiral field.


\section{Three-State Description of Helix, Coil, and Chirality}

We now introduce the minimal discrete model used to describe fluctuating
helical and chiral domains.  Consider a polymer chain consisting of $N$
coarse-grained segments.  Each segment is allowed to exist in one of three
local states, $ s_i=0,+1,-1$ ,
where $s_i=0$ denotes a coil-like segment, $s_i=+1$ denotes a right-handed
helical segment, and $s_i=-1$ denotes a left-handed helical segment.  The local
helical occupancy is $ n_i=s_i^2 $ .

Thus $n_i=1$ measures whether the segment is helical, whereas the sign of
$s_i$ specifies the handedness of the helical segment.  This separation between
occupancy and handedness is central to the present theory.  A chain may contain
locally helical segments without maintaining a single handedness over long
distances.  Conversely, once a helical segment has selected one handedness, it
may propagate that handedness over many segments if the cost of a handedness
reversal is large.

The three-state description is motivated by classical helix--coil theories,
particularly those of Zimm and Bragg, Gibbs and DiMarzio, and Lifson and Roig
\cite{ZimmBragg1958,GibbsDiMarzio1958,LifsonRoig1961}.  Those theories describe
the cooperative competition between a locally ordered helical state and a
disordered coil state.  The present formulation preserves that cooperative
helix--coil structure, but extends it by retaining the two possible helical
handedness states explicitly.  This allows one to describe not only the
formation of finite helical domains, but also the persistence, reversal, and
fluctuation of chirality within those domains.

A crucial ingredient is the entropy of the coil state.  A coil segment is not a
single microscopic configuration; it represents a large manifold of rotational,
torsional, and solvent-dependent conformations.  We therefore assign to the
coil state a local degeneracy factor
\begin{equation}
g=\exp\left(\frac{S_{\rm coil}}{k_B}\right),
\label{eq:coil_degeneracy}
\end{equation}
where $S_{\rm coil}$ is the local conformational entropy associated with the
coil state.  Equivalently, the coil state acquires an entropic free-energy
stabilization,
\begin{equation}
F_{\rm coil}\rightarrow F_{\rm coil}-k_BT\ln g .
\label{eq:coil_entropy_free_energy}
\end{equation}
This factor is essential.  Without coil degeneracy, the high-temperature limit
would assign comparable weights to the three local states $s_i=0,+1,-1$.
Since two of these three states are helical, the helical fraction would
incorrectly approach
\begin{equation}
\phi=\langle s_i^2\rangle \rightarrow \frac{2}{3}
\label{eq:wrong_highT_limit}
\end{equation}
at high temperature.  This is unphysical for a flexible polymer, where the coil
state should dominate at high temperature.  The degeneracy factor $g$ restores
the correct physical tendency by making the coil state entropically favored.

We next construct the minimal Hamiltonian.  The central point is that
cooperative stabilization should apply to a same-handed helical continuation,
but not to a direct contact between opposite-handed helices.  A neighboring
pair $+|-$ or $-|+$ should be regarded as a chirality-domain-wall defect, not
as a normally propagated helical bond.  This distinction is enforced by
introducing two bond projectors.

The same-handed helical propagation projector is
\begin{equation}
P_{\rm same}(i,i+1)
=
\frac{s_i^2s_{i+1}^2+s_is_{i+1}}{2}.
\label{eq:Psame}
\end{equation}
It is equal to unity for the two same-handed helical contacts,
$ +|+,\qquad -|- $,
and vanishes otherwise.  The opposite-handed or chiral-wall projector is
\begin{equation}
P_{\rm wall}(i,i+1)
=
\frac{s_i^2s_{i+1}^2-s_is_{i+1}}{2}.
\label{eq:Pwall}
\end{equation}
It is equal to unity for the two wrong-handed helical contacts,
$+|-,\qquad -|+$,
and vanishes otherwise.

The minimal Hamiltonian is then written as
\begin{equation}
H=
\sum_{i=1}^{N}\epsilon s_i^2
-
K\sum_{i=1}^{N-1}
\frac{s_i^2s_{i+1}^2+s_is_{i+1}}{2}
+
J\sum_{i=1}^{N-1}
\frac{s_i^2s_{i+1}^2-s_is_{i+1}}{2}
-
h\sum_{i=1}^{N}(1-s_i^2).
\label{eq:minimal_hamiltonian}
\end{equation}
The first term gives the effective local free-energy cost of forming a helical
unit.  As discussed in Sec.~2, this parameter includes bending, torsional,
hydrogen-bonding, packing, side-chain, and also solvent contributions.
The second term stabilizes same-handed helical propagation.  Thus $K>0$ is the
effective cooperative propagation free energy for a helical segment of fixed
handedness.  The third term penalizes a direct contact between helical segments
of opposite handedness.  Thus $J>0$ is the additional chiral-domain-wall
penalty.  The last term stabilizes the coil state; in practice, the coil
stabilization is determined by both the energetic parameter $h$ and the
entropic degeneracy factor $g$.

This form avoids an important ambiguity.  A chirality-independent term of the
form
\begin{equation}
-Ks_i^2s_{i+1}^2
\end{equation}
would stabilize any neighboring pair of helical segments, including the
wrong-handed contacts $+|-$ and $-|+$.  Such a term is inappropriate if a
handedness reversal requires a local defect, partial unwinding, loss of
torsional registry, disruption of hydrogen bonding, or packing frustration.
The projector form in Eq.~\eqref{eq:minimal_hamiltonian} instead assigns the
cooperative stabilization only to same-handed propagation and treats the
opposite-handed contact as a domain-wall excitation.

The pair energies implied by Eq.~\eqref{eq:minimal_hamiltonian} are
\begin{align}
E_{++}=E_{--} &= -K, 
\label{eq:pair_same_energy}\\
E_{+-}=E_{-+} &= +J .
\label{eq:pair_wall_energy}
\end{align}
Thus a wrong-handed contact loses the same-handed propagation stabilization
and also pays the explicit wall penalty.  The free-energy cost of replacing a
same-handed helical contact by a wrong-handed contact is therefore
\begin{equation}
\Delta F_{\rm wall}=K+J .
\label{eq:wall_free_energy}
\end{equation}
This quantity controls the rarity of handedness reversals, the density of
chirality domain walls, and the chiral correlation length.

The emergence of local homochirality in this model is therefore simple and
physical.  In the absence of an external chiral field, right- and left-handed
helical domains are degenerate.  A helical nucleus may choose either sign.  But
once a handedness has been chosen, propagation with the same handedness is
favored by $K$, while a reversal costs the wall free energy $K+J$.  Therefore
homochiral helical domains emerge not because one handedness is externally
preferred, but because mixed-handed domains contain energetically costly walls.

Eq. (25) possesses an important discrete symmetry. In the absence of an externally imposed chiral bias, the Hamiltonian is invariant under interchange of left- and right-handed helical states,
\begin{equation}
R \leftrightarrow L,
\qquad
C \rightarrow C .
\end{equation}

This transformation generates a $(Z_2)$ symmetry analogous to the spin-reversal symmetry of the Ising model. Although the local state space is three-dimensional, the chirality variable belongs to the odd representation of this symmetry, while the coil state remains invariant. As shown later through the transfer-matrix formulation, this symmetry naturally separates the spectrum into even and odd sectors and gives rise to an Ising-like form for the chirality correlation function and correlation length.

Although the Hamiltonian in Eq.~\eqref{eq:minimal_hamiltonian} is useful for
physical interpretation, the analytical theory is most transparent when written
directly in transfer-matrix form.  This is developed in the next section.

\section{Transfer Matrix and Symmetric--Antisymmetric Decomposition}
We now formulate the statistical mechanics of the three-state model mentioned in Sec.4, by using a
transfer matrix.  The transfer-matrix method is especially powerful in
one-dimensional Ising-like systems with nearest-neighbor interactions because
the full partition function can be written as a product of local statistical
weights.  The thermodynamic limit is then governed by the eigenvalue spectrum
of a finite-dimensional matrix.  The largest eigenvalue determines the free
energy per segment, while the subleading eigenvalues determine correlation
lengths.  The corresponding eigenvectors are not discarded: they determine the
composition of the thermodynamic state and the amplitudes with which different
observables couple to the various eigenmodes.  Thus the eigenvalues give the
length scales, while the eigenvectors determine which physical variables see
which length scales.

For a chain with nearest-neighbor interactions, the partition function can be
written schematically as
\begin{equation}
Z_N={\rm Tr}\,T^N ,
\label{eq:ZN_transfer_general}
\end{equation}
where $T$ is the transfer matrix.  If the eigenvalues of $T$ are
$\lambda_\alpha$, then
\begin{equation}
Z_N=\sum_\alpha \lambda_\alpha^N .
\label{eq:ZN_eigen_sum}
\end{equation}
For a positive transfer matrix, the Perron--Frobenius theorem ensures that the
largest eigenvalue is real, positive, and dominates the thermodynamic limit.
Denoting this eigenvalue by $\lambda_+$, one obtains
\begin{equation}
f=-k_BT\ln\lambda_+ .
\label{eq:free_energy_from_largest_lambda}
\end{equation}

The subleading eigenvalues do not affect the bulk free energy in the
thermodynamic limit, but they are essential for correlation functions.  A
typical two-point correlation function has the spectral form
\begin{equation}
\langle O_i O_{i+j}\rangle-\langle O\rangle^2
=
\sum_{\alpha\neq +}
{\cal A}_\alpha^{(O)}
\left(\frac{\lambda_\alpha}{\lambda_+}\right)^j ,
\label{eq:correlation_spectral_general}
\end{equation}
where the amplitudes ${\cal A}_\alpha^{(O)}$ are determined by matrix elements
of the operator $O$ between the dominant eigenvector and the subleading
eigenvectors.  Thus eigenvectors play a crucial role: if an observable has zero
overlap with a particular eigenmode by symmetry, that mode does not contribute
to the corresponding correlation function.  In the present problem this point
is central.  The helical occupancy $s_i^2$ is even under interchange of
right- and left-handed helices, whereas the chirality $s_i$ is odd.  As a
result, helical and chiral correlations couple to different sectors of the
transfer matrix.

The local states are ordered as $(-,0,+)$.
In this basis, the local helical-occupancy and chirality operators are
\begin{equation}
\widehat n =
\begin{pmatrix}
1&0&0\\
0&0&0\\
0&0&1
\end{pmatrix},
\qquad
\widehat s =
\begin{pmatrix}
-1&0&0\\
0&0&0\\
0&0&1
\end{pmatrix}.
\end{equation}
The occupancy operator $\widehat n$ is even under the interchange
$+\leftrightarrow-$, whereas the chirality operator $\widehat s$ is odd.
Consequently, $\widehat n$ couples the dominant symmetric eigenstate to
the subleading symmetric mode, while $\widehat s$ couples it to the
antisymmetric mode.

The transfer matrix has the compact form
\begin{equation}
T=
\begin{pmatrix}
A & M & B\\
M & C & M\\
B & M & A
\end{pmatrix}.
\label{eq:compact_transfer_matrix}
\end{equation}
The four statistical weights have the following meanings:
\begin{align}
A &= \text{same-handed helical continuation weight}, \nonumber\\
B &= \text{wrong-handed helical contact weight}, \nonumber\\
C &= \text{coil--coil continuation weight}, \nonumber\\
M &= \text{helix--coil boundary weight}.
\label{eq:weight_meanings}
\end{align}
The notation is deliberately general.  It allows the same transfer-matrix
solution to be used whether the microscopic origin of the helical stabilization
is hydrogen bonding, packing, side-chain interactions, torsional elasticity, or
a combination of these effects.

For the Hamiltonian in Eq.~\eqref{eq:minimal_hamiltonian}, with local terms
split symmetrically between neighboring bonds, a useful microscopic
identification is
\begin{align}
A &= \exp[-\beta\epsilon+\beta K],
\label{eq:A_weight}\\
B &= \exp[-\beta\epsilon-\beta J],
\label{eq:B_weight}\\
C &= g\exp(\beta h),
\label{eq:C_weight}\\
M &= \exp(-\beta\gamma_{hc})\,g^{1/2}
     \exp\left[-\frac{\beta}{2}(\epsilon-h)\right].
\label{eq:M_weight}
\end{align}
Here $\gamma_{hc}$ is the additional free-energy cost of a helix--coil boundary.
The factor $g^{1/2}$ in $M$ arises because a helix--coil bond contains one coil
state, while the factor $g$ in $C$ arises because a coil--coil bond contains the
full local coil degeneracy.  The parameter $M$ is therefore analogous to the
nucleation weight in the Zimm--Bragg description.  A small value of $M$ relative
to the geometric mean of the helix and coil weights corresponds to strong
cooperativity and a large cost of initiating or terminating a helical domain
\cite{ZimmBragg1958,LifsonRoig1961}.

The ratio of wrong-handed to same-handed helical weights is
\begin{equation}
\frac{B}{A}
=
\exp[-\beta(K+J)]
=
\exp[-\beta\Delta F_{\rm wall}].
\label{eq:B_over_A}
\end{equation}
Thus the wall free energy $\Delta F_{\rm wall}=K+J$ controls the rarity of
opposite-handed contacts.  
It replaces the simpler Ising-like parametrization in which the wall penalty is
often written as $2J$.

The special structure of Eq.~\eqref{eq:compact_transfer_matrix} follows from the
absence of an intrinsic chiral bias.  In the absence of an external chiral field,
right- and left-handed helical states are degenerate.  The transfer matrix is
therefore invariant under the interchange $+\leftrightarrow - $.
Equivalently, the transfer matrix commutes with the parity operation $P$ defined
by
\begin{equation}
P|+\rangle=|-\rangle,\qquad
P|-\rangle=|+\rangle,\qquad
P|0\rangle=|0\rangle .
\end{equation}
Since $T$ and $P$ commute, their eigenvectors may be chosen to have definite
parity under the exchange of right- and left-handed states.  This motivates the
introduction of the symmetric and antisymmetric helical combinations
\begin{equation}
|H_s\rangle=\frac{|+\rangle+|-\rangle}{\sqrt{2}},
\label{eq:Hs_def}
\end{equation}
and
\begin{equation}
|H_a\rangle=\frac{|+\rangle-|-\rangle}{\sqrt{2}}.
\label{eq:Ha_def}
\end{equation}
The coil state $|0\rangle$ is symmetric under $+\leftrightarrow -$.  Therefore
it can couple to $|H_s\rangle$, but it cannot couple to $|H_a\rangle$.  This is
the reason why the problem separates into a symmetric helix--coil sector and an
antisymmetric chirality sector.

This separation can be shown explicitly.  Acting on the antisymmetric state,
\begin{equation}
T|H_a\rangle
=
T\frac{|+\rangle-|-\rangle}{\sqrt{2}} .
\end{equation}
Since $|+\rangle$ and $|-\rangle$ couple equally to the coil state, the coil
contributions cancel.  The only remaining contribution is the difference between
same-handed and wrong-handed helical contacts:
\begin{equation}
T|H_a\rangle=(A-B)|H_a\rangle .
\label{eq:antisym_eigen}
\end{equation}
Thus the antisymmetric state is already an eigenvector, with eigenvalue
\begin{equation}
\lambda_\chi=A-B .
\label{eq:lambda_chi}
\end{equation}
This eigenvalue controls the decay of handedness correlations.

In the transformed basis
\begin{equation}
(|0\rangle,|H_s\rangle,|H_a\rangle),
\end{equation}
the transfer matrix becomes block diagonal:
\begin{equation}
T\rightarrow
\begin{pmatrix}
C & \sqrt{2}M & 0\\
\sqrt{2}M & A+B & 0\\
0 & 0 & A-B
\end{pmatrix}.
\label{eq:block_diagonal_matrix}
\end{equation}
The two-dimensional block describes the competition between coil and helical
occupancy. The one-dimensional antisymmetric block describes handedness fluctuations and the decay of chiral correlations.

The two eigenvalues of the symmetric block are
\begin{equation}
\lambda_{\pm}
=
\frac{1}{2}
\left[
C+A+B
\pm
\sqrt{(C-A-B)^2+8M^2}
\right],
\label{eq:lambda_pm}
\end{equation}
while the antisymmetric eigenvalue is
\begin{equation}
\lambda_\chi=A-B.
\label{eq:lambda_chi_again}
\end{equation}
For a long chain, the partition function behaves as
\begin{equation}
Z_N\simeq \lambda_+^N,
\label{eq:partition_largeN}
\end{equation}
where $\lambda_+$ is the largest eigenvalue.  The free energy per segment is
therefore
\begin{equation}
f=-k_BT\ln\lambda_+ .
\label{eq:free_energy_segment}
\end{equation}

The eigenvectors also provide physical information.  The dominant eigenvector
associated with $\lambda_+$ gives the equilibrium mixture of coil and helical
components in the thermodynamic state.  In the symmetric block, its components
determine the relative weight of the coil state $|0\rangle$ and the symmetric
helical combination $|H_s\rangle$.  The subleading symmetric eigenvector
associated with $\lambda_-$ controls fluctuations of helical occupancy.  The
antisymmetric eigenvector $|H_a\rangle$ controls fluctuations of handedness.
Thus the eigenfunctions are not auxiliary mathematical objects: they determine
which physical observables couple to which eigenvalues.

The dominant symmetric eigenvector also gives the equilibrium helical
fraction. Writing
\begin{equation}
D=\sqrt{(C-A-B)^2+8M^2},
\end{equation}
one obtains
\begin{equation}
\phi_H\equiv \langle s_i^2\rangle
=
\frac{1}{2}
\left[
1-\frac{C-A-B}{D}
\right].
\end{equation}
The corresponding coil fraction is $1-\phi_H$. Thus the dominant
eigenvector determines the equilibrium composition, while the
subleading eigenvalues determine the spatial decay of fluctuations.

It is useful to note that the corrected weights still retain an Ising-like
hyperbolic-function structure.  From Eqs.~\eqref{eq:A_weight} and
\eqref{eq:B_weight},
\begin{align}
A+B
&=
2\exp\left[-\beta\epsilon+\frac{\beta}{2}(K-J)\right]
\cosh\left[\frac{\beta}{2}(K+J)\right],
\label{eq:AplusB_corrected}\\
A-B
&=
2\exp\left[-\beta\epsilon+\frac{\beta}{2}(K-J)\right]
\sinh\left[\frac{\beta}{2}(K+J)\right].
\label{eq:AminusB_corrected}
\end{align}
Thus the relevant chiral-wall free energy is $K+J$, not $2J$.  In the
special symmetric Ising parametrization one may set $\Delta F_{\rm wall}=2J$,
but in the present polymer problem the more physical result is
\begin{equation}
\Delta F_{\rm wall}=K+J.
\label{eq:wall_energy_summary}
\end{equation}

The block diagonalization in Eq.~\eqref{eq:block_diagonal_matrix} is central to
the present theory.  It shows that helix--coil thermodynamics and chirality
fluctuations are coupled through the same transfer matrix, but are governed by
different eigenmodes.  The symmetric sector controls the helix fraction and
helical-domain length, whereas the antisymmetric sector controls the persistence of handedness.

\section{Helix Breaking and Handedness Reversal:
Two Cooperative Length Scales}

A finite chiral helix can lose its order in two physically distinct
ways. First, the chain may leave the helical manifold and enter the
coil state, thereby terminating the helical segment. Second, the chain
may remain helical but reverse its handedness, thereby creating a
chiral domain wall. These two processes are controlled by different
defects and therefore generate different cooperative length scales.

The first is the helix--coil correlation length $\xi_H$, which measures
the persistence of helical occupancy and is controlled primarily by
the cost of creating helix--coil boundaries. The second is the chiral
persistence length $\xi_\chi$, which measures the persistence of one
handedness within the helical manifold and is controlled by the
free-energy cost of a handedness-reversal wall. A physically observed
homochiral segment is limited by whichever of these two mechanisms
destroys its order first.

This distinction follows directly from symmetry. The helical occupancy
$n_i=s_i^2$ is even under $+\leftrightarrow-$, whereas the handedness
$s_i$ is odd. Consequently, occupancy fluctuations couple to the
subleading symmetric eigenvalue $\lambda_-$, while handedness
fluctuations couple to the antisymmetric eigenvalue
$\lambda_\chi=A-B$.

The correlation lengths derived below follow directly from the
spectral representation of the transfer matrix. For an observable
$O$, the connected two-point correlation function can be written as a
sum of contributions proportional to
$(\lambda_\alpha/\lambda_+)^j$, where $\lambda_+$ is the dominant
eigenvalue and $\lambda_\alpha$ denotes a subleading eigenvalue. At
separations beyond the short local-transient regime, the slowest
symmetry-allowed subleading mode dominates. The correlation function
then decays exponentially, and the associated correlation length is
given by
\begin{equation}
\xi^{-1}
=
\ln\!\left(
\frac{\lambda_+}{|\lambda_{\rm sub}|}
\right).
\end{equation}

Equation (59) provides the general spectral relation; we now apply it separately to helical-occupancy and handedness correlations.


\subsection{Helix--Coil Correlation Length}

The connected helical correlation function is defined as
\begin{equation}
G_H(j)
=
\langle n_i n_{i+j}\rangle-\langle n_i\rangle^2 .
\label{eq:helix_corr_def}
\end{equation}
For a one-dimensional transfer-matrix system with a discrete gapped spectrum,
this correlation function decays exponentially at large separation:
\begin{equation}
G_H(j)\sim \exp(-j/\xi_H).
\label{eq:helix_corr_decay}
\end{equation}
Since $n_i$ is even under chirality reversal, the leading contribution to
$G_H(j)$ comes from the two eigenvalues of the symmetric helix--coil sector.
Thus
\begin{equation}
\xi_H^{-1}
=
\ln\left(\frac{\lambda_+}{|\lambda_-|}\right).
\label{eq:xiH_general}
\end{equation}
Using Eq.~\eqref{eq:lambda_pm}, this gives
\begin{equation}
\xi_H^{-1}
=
\ln
\left[
\frac{
C+A+B+\sqrt{(C-A-B)^2+8M^2}
}{
C+A+B-\sqrt{(C-A-B)^2+8M^2}
}
\right].
\label{eq:xiH_explicit}
\end{equation}

Near the helix--coil crossover, the coil and helical weights are comparable:
\begin{equation}
C\simeq A+B .
\label{eq:crossover_condition}
\end{equation}
Then
\begin{equation}
\lambda_\pm\simeq C\pm \sqrt{2}M .
\label{eq:lambda_pm_crossover}
\end{equation}
If the helix--coil boundary weight is written in Zimm--Bragg form as
\begin{equation}
M=\sqrt{\sigma}\,C
\label{eq:M_sigma}
\end{equation}
at coexistence, where $\sigma$ is the nucleation parameter, then the helical
correlation length scales as
\begin{equation}
\xi_H\sim \sigma^{-1/2}.
\label{eq:xiH_sigma_scaling}
\end{equation}
This is the familiar cooperative-length scaling of the Zimm--Bragg theory,
recovered here in a three-state chiral model.  Thus $\xi_H$ gives the typical
size of a contiguous helical domain, irrespective of whether that domain is
uniformly right-handed or left-handed.

\subsection{Chiral Persistence Length and Domain Walls}

We next consider chirality.  The chiral correlation function is given by

\begin{equation}
C_\chi(j)=\langle s_i s_{i+j}\rangle .
\label{eq:chiral_corr_def}
\end{equation}
Because $s_i$ is odd under $+\leftrightarrow -$, this correlation function
couples to the antisymmetric eigenmode.  Therefore
\begin{equation}
C_\chi(j)\sim \exp(-j/\xi_\chi),
\label{eq:chiral_corr_decay}
\end{equation}
with
\begin{equation}
\xi_\chi^{-1}
=
\ln\left(\frac{\lambda_+}{|\lambda_\chi|}\right).
\label{eq:xichi_general}
\end{equation}
Using $\lambda_\chi=A-B$, one obtains
\begin{equation}
\xi_\chi^{-1}
=
\ln\left(\frac{\lambda_+}{|A-B|}\right).
\label{eq:xichi_explicit}
\end{equation}
This length measures the persistence of handedness.  It is therefore distinct
from the helical occupancy length $\xi_H$.

The corrected Hamiltonian gives the same-handed and wrong-handed helical
weights as
\begin{equation}
A=\exp[-\beta\epsilon+\beta K],
\label{eq:A_sec3}
\end{equation}
and
\begin{equation}
B=\exp[-\beta\epsilon-\beta J].
\label{eq:B_sec3}
\end{equation}
Thus the ratio of wrong-handed to same-handed helical contacts is
\begin{equation}
\frac{B}{A}
=
\exp[-\beta(K+J)] .
\label{eq:B_over_A_sec3}
\end{equation}
The relevant domain-wall free energy is therefore
\begin{equation}
\Delta F_{\rm wall}=K+J.
\label{eq:wall_free_energy_sec3}
\end{equation}
This quantity has a simple physical meaning.  A wrong-handed contact loses the
same-handed propagation stabilization $K$ and also pays the explicit
chiral-wall penalty $J$.

The hyperbolic functions still enter naturally, but their argument is now the
full wall free energy.  From Eqs.~\eqref{eq:A_sec3} and \eqref{eq:B_sec3},
\begin{equation}
A+B
=
2\exp\left[-\beta\epsilon+\frac{\beta}{2}(K-J)\right]
\cosh\left[\frac{\beta}{2}(K+J)\right],
\label{eq:AplusB_corrected}
\end{equation}
and
\begin{equation}
A-B
=
2\exp\left[-\beta\epsilon+\frac{\beta}{2}(K-J)\right]
\sinh\left[\frac{\beta}{2}(K+J)\right].
\label{eq:AminusB_corrected}
\end{equation}
Thus the even helix-forming sector is controlled by the $\cosh$ combination,
whereas the odd chiral sector is controlled by the $\sinh$ combination.  The
important difference from the simpler Ising parametrization is that the
argument is $\beta(K+J)/2$, not $\beta J$.

In the strongly helical limit, coil interruptions are rare and
\begin{equation}
\lambda_+\simeq A+B .
\label{eq:strong_helix_lambda}
\end{equation}
Then
\begin{equation}
\xi_\chi^{-1}
\simeq
\ln\left(\frac{A+B}{A-B}\right)
=
\ln
\left[
\coth\left(\frac{\beta(K+J)}{2}\right)
\right].
\label{eq:xichi_coth_corrected}
\end{equation}
For a large wall free energy, $\beta(K+J)\gg 1$,
\begin{equation}
\coth\left(\frac{\beta(K+J)}{2}\right)
\simeq
1+2e^{-\beta(K+J)} ,
\end{equation}
and therefore
\begin{equation}
\xi_\chi
\simeq
\frac{1}{2}e^{\beta(K+J)}.
\label{eq:xichi_corrected_scaling}
\end{equation}
Equivalently,
\begin{equation}
\xi_\chi
\simeq
\frac{1}{2}e^{\beta\Delta F_{\rm wall}} .
\label{eq:xichi_wall_scaling}
\end{equation}
This is the corrected Ising-like result for the present polymer model.  The
older form $\xi_\chi\simeq (1/2)e^{2\beta J}$ is recovered only in the special
parametrization $\Delta F_{\rm wall}=2J$.  In the present theory, the physical
wall penalty is instead $K+J$.

The two correlation lengths therefore have different origins:
\begin{equation}
\xi_H\sim \sigma^{-1/2},
\qquad
\xi_\chi\sim \frac{1}{2}e^{\beta(K+J)}.
\label{eq:two_lengths_summary_corrected}
\end{equation}
The first length is controlled by helix--coil nucleation and cooperativity.  The
second is controlled by the free-energy cost of a local handedness reversal.
This separation is one of the main results of the theory.

The same physics can be expressed in terms of chirality domain walls.  A chiral
wall is a neighboring pair of helical segments with opposite handedness,
\begin{equation}
+|- \qquad \text{or} \qquad -|+ .
\end{equation}
The relative statistical weight of an opposite-handed helical contact compared
with a same-handed contact is
\begin{equation}
\frac{B}{A}
=
e^{-\beta(K+J)}
=
e^{-\beta\Delta F_{\rm wall}} .
\label{eq:wall_weight_ratio_corrected}
\end{equation}
The probability of a chiral wall inside a helical region is therefore
approximately
\begin{equation}
p_{\rm wall}
=
\frac{B}{A+B}
=
\frac{1}{1+e^{\beta(K+J)}} .
\label{eq:wall_probability_corrected}
\end{equation}
For large $\beta(K+J)$,
\begin{equation}
p_{\rm wall}\simeq e^{-\beta(K+J)}
=
e^{-\beta\Delta F_{\rm wall}} .
\label{eq:wall_probability_largeKJ}
\end{equation}
The average distance between chiral walls is then
\begin{equation}
L_{\rm wall}
\sim p_{\rm wall}^{-1}
\sim e^{\beta(K+J)} .
\label{eq:domain_size_wall_corrected}
\end{equation}
This is consistent with Eq.~\eqref{eq:xichi_corrected_scaling}, apart from a
factor of order unity.  In the rare-wall limit, the sign correlation length is
approximately one half of the mean distance between walls:
\begin{equation}
\xi_\chi\simeq \frac{1}{2}L_{\rm wall}.
\label{eq:xichi_Lwall_relation}
\end{equation}

This domain-wall interpretation also clarifies the origin of local
homochirality.  In the absence of an external chiral field, right- and
left-handed helical domains are degenerate.  The theory does not favor one sign
globally.  However, a mixed-handed helical domain contains walls, and each wall
costs $\Delta F_{\rm wall}=K+J$.  Therefore a finite helical domain tends to
remain homochiral over a length set by $\xi_\chi$.  Homochirality here is thus
a local domain property generated by the suppression of energetically costly
handedness reversals.


In a real polymer chain, the observable chiral domain size is limited by both
the spatial extent of the helical segment and the intrinsic handedness
persistence length.  A helical segment cannot maintain chirality beyond the
length over which it remains helical.  If helix--coil interruptions and chiral
walls are treated as independent sources of decorrelation, a useful estimate is
\begin{equation}
\frac{1}{L_{\rm obs}}
\simeq
\frac{1}{\xi_H}
+
\frac{1}{\xi_\chi}.
\label{eq:Lobs_inverse}
\end{equation}
Thus
\begin{equation}
L_{\rm obs}
\simeq
\left(
\xi_H^{-1}+\xi_\chi^{-1}
\right)^{-1}.
\label{eq:Lobs}
\end{equation}
Equivalently, up to factors of order unity,
\begin{equation}
L_{\rm obs}\sim \min(\xi_H,\xi_\chi).
\label{eq:Lobs_min}
\end{equation}
This result is useful experimentally because the measured chiral persistence
may be controlled either by finite helix size or by genuine handedness
reversals.

The transfer matrix gives equilibrium correlation lengths, but experiments
often probe lifetimes.  To estimate lifetimes, we introduce a simple kinetic
closure.  Consider a finite helical domain of length $L$.  Let $k_+$ be the rate
for adding a helical unit at a domain edge, and let $k_-$ be the rate for losing
a helical unit at a domain edge.  Away from coexistence, where the drift of the
domain boundary is nonzero, the mean lifetime of a helical domain is
approximately
\begin{equation}
\tau_H(L)
\simeq
\frac{L}{2(k_- - k_+)} .
\label{eq:tauH_drift}
\end{equation}
The factor of two appears because a finite helix has two fraying ends.  Near
coexistence, where $k_+\simeq k_-$, the boundary motion becomes diffusive in
domain-size space.  Then
\begin{equation}
\tau_H(L)
\simeq
\frac{L^2}{2(k_+ + k_-)} .
\label{eq:tauH_diffusive}
\end{equation}
Since the typical equilibrium domain length is $L\sim \xi_H$, the near-crossover
lifetime scales as
\begin{equation}
\tau_H\sim \xi_H^2\sim \sigma^{-1}.
\label{eq:tauH_sigma}
\end{equation}
This relation should be understood as a kinetic estimate based on end fraying,
not as an exact equilibrium result.


\section{Application to Protein $\alpha$-Helices:
Domain Lengths and Wall Free Energies}


The analytical results derived above can be used in reverse. Rather
than choosing the model parameters first and calculating a domain
length, one may begin with an experimentally observed helical length
and ask what range of the cooperativity parameter and chiral-wall free
energy would be consistent with it. We apply this procedure to three
representative protein-like systems: ordinary short $\alpha$-helices
in globular proteins, alanine-based helical peptides, and a long
single-$\alpha$-helical domain of myosin VI.

The estimates made below are intended to establish physical scales.
They are not residue-specific fits, because an observed helical length
need not be identical to either intrinsic correlation length. In
particular, a homochiral domain may terminate either through a
helix--coil boundary or through a handedness-reversal wall.

The transfer-matrix parameters introduced  summarize, at a coarse-grained level, the combined effects
of bending, torsion, hydrogen bonding, packing, side-chain interactions, and
solvent.  This interpretation is essential if the theory is to be connected to experiments or simulations. 

The parameters \(\epsilon\), \(K\), and \(J\) are coarse-grained free energies containing the physical contributions discussed in Sec.~3. Here we estimate only the combinations constrained by observed
helical and homochiral lengths.

\subsection{From an Observed Helix Length to Model Parameters}

If a measured contiguous helical length $L_H$ is identified, at the
level of an order-of-magnitude estimate, with the helix--coil
correlation length, then
\begin{equation}
\sigma_{\rm est}\sim L_H^{-2}.
\end{equation}
Likewise, if an observed homochiral length $L_\chi$ is identified with
the intrinsic chiral persistence length, then
\begin{equation}
\frac{\Delta F_{\rm wall}}{k_BT}
\simeq \ln(2L_\chi).
\end{equation}
Under these stated identifications, the two inverse relations provide
order-of-magnitude estimates of the cooperativity parameter and the
chiral-wall free energy.


\subsection{Ordinary Protein $\alpha$-Helices}

Structural surveys indicate that an ordinary $\alpha$-helix in a
globular protein frequently contains of order ten residues. Taking
$L_H\simeq10$ as a representative contiguous helical length gives
\begin{equation}
\sigma_{\rm est}\sim10^{-2}.
\end{equation}

If this ten-residue interval is also identified, only as an
order-of-magnitude estimate, with the handedness-persistence length,
the corresponding wall free energy is
\begin{equation}
\frac{\Delta F_{\rm wall}}{k_BT}
\simeq\ln(20)\simeq3.0.
\end{equation}

Thus a relatively short protein helix is consistent with a moderate
helix--coil cooperativity and a chiral-wall free energy of only a few
$k_BT$. The short observed length does not imply weak local
handedness stabilization: the domain may terminate through a
helix--coil boundary before a handedness-reversal wall is encountered.

\subsection{Alanine-Based Peptides and Long Single Helices}

Alanine-based helix--coil experiments commonly employ chains in the
range of approximately 19--39 residues. Taking representative lengths
of 20 and 40 residues gives
\begin{equation}
\sigma_{\rm est}
\sim 2.5\times10^{-3}
\quad\text{and}\quad
6.3\times10^{-4},
\end{equation}
respectively. If the entire helical segment retains a common
handedness, the corresponding wall free energies are
\begin{equation}
\frac{\Delta F_{\rm wall}}{k_BT}
\simeq 3.69
\quad\text{and}\quad
4.38.
\end{equation}

A longer example is provided by the medial tail domain of myosin VI,
where an ordered single helix extends over roughly 58 residues.
Identifying this length, only for the purpose of estimating scale,
with the relevant persistence length gives
\begin{equation}
\sigma_{\rm est}\sim
\frac{1}{58^2}
\simeq 3.0\times10^{-4},
\end{equation}
and
\begin{equation}
\frac{\Delta F_{\rm wall}}{k_BT}
\simeq \ln(116)
\simeq 4.75.
\end{equation}

The striking result is that increasing the persistence from about ten
to nearly sixty residues requires the wall free energy to increase
only from approximately $3\,k_BT$ to approximately $4.8\,k_BT$.
This modest energetic change produces a large increase in domain
length because the chiral persistence depends exponentially on the
wall free energy.

\subsection{Which Mechanism Limits the Observed Domain?}

The numerical estimates also show that the observable homochiral length
is not determined by the chiral-wall free energy alone. Recall that
\(\sigma=\exp(-2\beta\gamma_{hc})\) measures the cost of creating the two
helix--coil boundaries of a finite helical domain. Thus a small but
nonzero \(\sigma\) suppresses helix termination without eliminating it.
For example, \(\sigma=10^{-2}\) represents appreciable helix--coil
cooperativity, but still gives only
\(\xi_H\sim\sigma^{-1/2}\simeq10\) segments. By contrast,
\(\Delta F_{\rm wall}=5k_BT\) gives
\(\xi_\chi\simeq \tfrac12 e^5\simeq74\) segments. Handedness would
therefore persist over a much longer distance if the helix remained
intact, but the finite probability of forming a helix--coil boundary
terminates the helical segment first. The observable homochiral length
is consequently controlled mainly by \(\xi_H\), rather than by
\(\xi_\chi\), in this example.

Conversely, if $\sigma=10^{-4}$, so that $\xi_H$ is of order one
hundred segments, but $\Delta F_{\rm wall}=3k_BT$, so that
$\xi_\chi\simeq10$ segments, the chain may remain helical over a long
distance while its handedness reverses repeatedly. The observed
homochiral length is then limited by chiral walls rather than by
helix--coil boundaries.

This competition is summarized by
\begin{equation}
L_{\rm obs}
\simeq
\left(\xi_H^{-1}+\xi_\chi^{-1}\right)^{-1}.
\end{equation}
Long homochiral helices therefore require both strong helix--coil
cooperativity and a sufficiently large wall free energy.


\section{Dynamics of Chiral Walls and Helical Domains}
The transfer matrix determines equilibrium probabilities and correlation
lengths, but it does not by itself specify the microscopic transition
rates between local conformations. We therefore introduce a minimal
kinetic description based on the physical defects identified above.
The resulting expressions are scaling estimates: they assume local
activated rearrangements and approximately diffusive motion of
helix--coil boundaries or chiral walls. No specific kinetic Ising
dynamics is imposed.

\subsection{Helix Survival and End Fraying}

A finite helical domain has two fluctuating boundaries. At either end,
a coil segment may become helical or a terminal helical segment may
unwind into the coil state. Away from the helix--coil crossover these
two processes are unbalanced. If unwinding is favored, the two
boundaries drift inward and the characteristic lifetime of a domain
increases approximately linearly with its initial length. If helical
growth is favored, the domain instead tends to expand until it meets
another domain, a chain end, or some structural constraint.

Near the helix--coil crossover, growth and loss occur at nearly equal
rates. The systematic drift of each boundary then becomes small, and
the domain length undergoes an approximately diffusive fluctuation.
The characteristic time required for a domain of length $L$ to change
by an amount comparable to its own size consequently grows as $L^2$.
This should be regarded as a diffusive relaxation time rather than as
a universal exact mean first-passage lifetime, because the latter
depends on chain length, boundary conditions, and the free-energy
variation with domain size. For a typical equilibrium domain with
$L\sim\xi_H$, the resulting scaling is
\begin{equation}
\tau_H^{\rm char}\sim \tau_0\,\xi_H^2
\sim \tau_0\,\sigma^{-1},
\end{equation}
where $\tau_0$ is a microscopic time associated with local
helix--coil conversion at a domain boundary. Thus strong
helix--coil cooperativity increases not only the spatial extent of a
helical domain but also its characteristic survival time.

\subsection{Chiral-Wall Nucleation and Propagation}

Chiral memory may be lost by a different route. Even when the chain
remains helical, a local handedness-reversal defect may be created
inside the domain. The nucleation of such a wall requires the
free-energy cost
$\Delta F_{\rm wall}=K+J$, and is therefore activated. If $k_0$ is a
local attempt frequency, the wall-creation rate per possible
nucleation site is of order
\begin{equation}
k_{\rm nuc}\sim
k_0\exp(-\beta\Delta F_{\rm wall}).
\end{equation}
For a helical domain containing $L$ possible nucleation sites, the
total creation rate is correspondingly of order
$Lk_{\rm nuc}$, provided the sites may be treated as approximately
independent. This length dependence is important: a longer helix
contains more possible locations at which a rare wall can be created.

Wall creation alone does not necessarily erase the handedness of the
entire domain. Once formed, the wall separates right- and left-handed
helical regions and may move along the chain through successive local
rearrangements. A one-segment displacement of the wall requires a
local change of handedness near the defect, possibly accompanied by
partial unwinding and reformation of the helical registry. If
$k_{\rm flip}$ denotes the characteristic rate for such a local wall
step and $b$ is the coarse-grained segment spacing, the wall diffusion
coefficient scales as
\begin{equation}
D_{\rm wall}\sim b^2 k_{\rm flip}.
\end{equation}
The precise numerical prefactor depends on whether $k_{\rm flip}$ is
defined as the rate in each direction or as the total rate for either
direction.

For an unbiased wall moving within a helical domain of length $L$, the
characteristic time required to explore or traverse the domain is
therefore
\begin{equation}
\tau_{\rm traverse}\sim
\frac{L^2}{D_{\rm wall}}.
\end{equation}
This introduces a second dynamical stage after nucleation. The total
time for reversal of a finite domain may be controlled either by the
activated waiting time for creation of the first wall or by the time
required for the wall to propagate across the domain. In a
nucleation-limited regime,
\begin{equation}
\tau_{\rm reverse}
\sim
\frac{\exp(\beta\Delta F_{\rm wall})}{Lk_0},
\end{equation}
whereas in a propagation-limited regime,
\begin{equation}
\tau_{\rm reverse}
\sim
\frac{L^2}{D_{\rm wall}}.
\end{equation}
More generally, the two contributions occur sequentially and the
reversal time may be estimated as
\begin{equation}
\tau_{\rm reverse}
\sim
\frac{\exp(\beta\Delta F_{\rm wall})}{Lk_0}
+
\frac{L^2}{D_{\rm wall}}.
\end{equation}
This expression is not an exact result, but it makes explicit the
competition between activated defect creation and diffusive defect
motion.

\subsection{Migration of an Intact Helical Domain}

A finite helix can also migrate along the polymer without changing its
length appreciably. Such migration occurs when one end grows while
the opposite end frays. Repetition of these coordinated end events
translates the helical segment along the sequence. If the two ends
fluctuate nearly independently and the rates for addition and loss
are comparable, the center of the helical domain performs an
approximately diffusive motion. Its diffusion coefficient is of the
form
\begin{equation}
D_H\sim b^2 k_{\rm end},
\end{equation}
where $k_{\rm end}$ is an effective rate for the paired end
rearrangements that shift the domain by one segment. The actual rate
can be considerably smaller than a single-end fraying rate because a
net translation requires correlated changes at opposite boundaries.

The dynamics therefore contains several physically distinct time
scales. The survival of helical occupancy is controlled mainly by
motion of the helix--coil boundaries and hence by the cooperativity
parameter $\sigma$. Loss of handedness requires either destruction of
the helical domain or activated creation and subsequent motion of a
chiral wall. Translation of an intact helix is governed by coordinated
growth and fraying at its ends. These processes need not occur on the
same time scale, even when they involve the same finite helical
segment.

This separation suggests direct experimental and simulation tests.
Time-resolved helix-sensitive probes measure the survival of helical
occupancy, whereas chiral spectroscopies probe persistence of
handedness. Single-molecule trajectories or residue-resolved
simulations may distinguish end fraying from internal wall creation
and may reveal whether a handedness reversal is nucleation-limited or
propagation-limited. The theory therefore predicts not one universal
helix lifetime, but a hierarchy of kinetic processes associated with
helix survival, chiral-wall formation, wall migration, and translation
of the complete helical domain.

\section{Intrinsic Chiral Bias and Right-Handed Helix Selection}
\label{sec:intrinsic_chiral_bias}

The purpose of this section is to explain how molecular chirality selects one
of the two otherwise symmetry-related helical senses and how that local
preference is amplified over a finite cooperative domain. For proteins, the
stereochemistry of L-amino-acid residues makes the right-handed
\(\alpha\)-helical basin of the Ramachandran plot shown in Fig.~2 lower in
free energy than the corresponding left-handed basin. Poly-L-alanine provides
a particularly simple molecular example of this intrinsic right-handed
preference, whereas synthetic helical polymers provide complementary examples
in which a weak local chiral bias is amplified over many repeat units. We first
introduce this intrinsic bias into the transfer-matrix theory and then apply
the resulting finite-domain amplification factor to proteins, polyalanine, and
synthetic helical polymers.

The present theory distinguishes two physically separate questions.  The
L-amino-acid backbone supplies a local stereochemical bias that selects the
right-handed $\alpha$-helical state, while the chiral domain-wall free energy
suppresses reversals and converts this local bias into persistent handedness
over many residues.  The local bias therefore determines which handedness is
preferred, whereas the domain-wall free energy determines how long that
handedness persists along the chain.

\subsection{Stereochemical Origin of the Intrinsic Chiral Bias}

In many synthetic or coarse-grained polymer models, the right- and left-handed
helical states may initially be treated as degenerate.  This is the situation
described in the preceding sections.  A helical nucleus can choose either
handedness, but after that choice has been made, a reversal creates a domain
wall.  The wall free energy then produces local homochirality.  Proteins
present a more specific problem.  Natural proteins are built almost entirely
from L-amino acids, and their $\alpha$-helices are overwhelmingly
right-handed.  Thus one must explain not only why a helical domain remains
homochiral, but also why the preferred homochiral state has a particular sign.

The driving force for this selection is local stereochemistry.  In an
L-amino-acid residue, the tetrahedral arrangement around the $C_{\alpha}$ atom
fixes the relative positions of the side chain, the amide group, and the
carbonyl group.  Consequently, the backbone torsion angles $(\phi,\psi)$ do
not explore all geometrically possible values with equal probability.  The
Ramachandran plot provides the standard representation of these steric
restrictions
\cite{Ramachandran1963,BrandenTooze,Creighton1993,Hovmoller2002}.
As illustrated in Fig.~2, for ordinary L-amino acids the right-handed
$\alpha$-helical basin lies in a more favored region of the allowed
$(\phi,\psi)$ space.  The corresponding left-handed $\alpha$-helical region
is much more restricted and is populated mainly under special local
circumstances, particularly by glycine and near turns or helix termini
\cite{Hovmoller2002,ShepherdFairlie2009,GunasekaranBalaram1998}.
The symbol L in ``L-amino acid'' denotes the molecular stereochemical
configuration of the residue; it does not denote the handedness of the
resulting helix.  The stereochemistry of L-amino acids instead favors the
right-handed $\alpha$-helix.

In the present notation, the local free energy of a right-handed helical
segment is therefore lower than that of a left-handed helical segment.  If
$s_i=+1$ denotes a right-handed helical state and $s_i=-1$ denotes a
left-handed helical state, this intrinsic preference may be represented by
adding the chiral-bias term
\begin{equation}
H_{\rm bias}
=
-\Delta_{\chi}\sum_i s_i .
\label{eq:chiral_bias_term}
\end{equation}

Equation~\eqref{eq:chiral_bias_term} is the analogue of the magnetic-field
term $-H\sum_i S_i$ in a ferromagnet.  The handedness variable $s_i$ plays
the role of an Ising spin, while $\Delta_\chi$ acts as a local chiral field.
The analogy is formal but physically useful: the field selects one sign,
whereas the cooperative interaction and the domain-wall free energy determine
how strongly that choice is propagated along the chain.  Unlike an externally
applied magnetic field, however, $\Delta_\chi$ is an intrinsic molecular field
generated by the stereochemistry of the L-amino-acid backbone.  It represents
an effective local free-energy difference between right- and left-handed
helical conformations.

For an L-amino-acid protein backbone, $\Delta_{\chi}>0$ with the convention
used here.  The parameter $\Delta_{\chi}$ summarizes the preference encoded
in the Ramachandran landscape, side-chain--backbone steric interactions,
hydrogen-bond geometry, local packing constraints, and solvent-dependent
effects.  Equivalently, one may write
\begin{equation}
\Delta_{\chi}
=
\frac{1}{2}
\left(
f_L^{\rm helix}
-
f_R^{\rm helix}
\right),
\label{eq:Delta_chi_definition}
\end{equation}
where $f_R^{\rm helix}$ and $f_L^{\rm helix}$ are the local free energies of
right- and left-handed helical residues, respectively.  For L-amino acids,
$f_R^{\rm helix}<f_L^{\rm helix}$ and therefore $\Delta_{\chi}>0$.  For a
chain made of D-amino acids, the stereochemical relation is reversed and the
corresponding chiral bias has the opposite sign.

The biased Hamiltonian is therefore
\begin{equation}
H_{\rm total}
=
H
-
\Delta_{\chi}\sum_i s_i ,
\label{eq:total_H_with_bias}
\end{equation}
where $H$ is the symmetric Hamiltonian of
Eq.~\eqref{eq:minimal_hamiltonian}.  The added term breaks the
$+\leftrightarrow-$ symmetry.  It does not replace the domain-wall physics;
rather, it selects which of the two homochiral states is preferred.

\subsection{Biased Transfer Matrix and Symmetry Breaking}

The corresponding transfer matrix is no longer symmetric between right- and
left-handed helices.  The same-handed propagation weights become
\begin{equation}
A_+
=
\exp[-\beta\epsilon+\beta K+\beta\Delta_{\chi}],
\label{eq:A_plus_bias}
\end{equation}
and
\begin{equation}
A_-
=
\exp[-\beta\epsilon+\beta K-\beta\Delta_{\chi}],
\label{eq:A_minus_bias}
\end{equation}
where $A_+$ corresponds to right-handed propagation and $A_-$ to
left-handed propagation.  The helix--coil boundary weights are similarly
modified:
\begin{equation}
M_+
=
M\exp\left(\frac{\beta\Delta_{\chi}}{2}\right),
\qquad
M_-
=
M\exp\left(-\frac{\beta\Delta_{\chi}}{2}\right).
\label{eq:M_pm_bias}
\end{equation}

A wrong-handed contact contains one right-handed and one left-handed segment.
The one-body chiral contributions therefore cancel in this bond, leaving the
wrong-handed contact weight
\begin{equation}
B
=
\exp[-\beta\epsilon-\beta J].
\label{eq:B_bias}
\end{equation}
Thus the local field selects the preferred helical sense, whereas the
wrong-handed contact remains controlled by the explicit mismatch penalty $J$
and by the loss of the same-handed propagation stabilization $K$.

In the ordered basis $(-,0,+)$, the biased transfer matrix may be written as
\begin{equation}
T_{\chi}
=
\begin{pmatrix}
A_- & M_- & B\\
M_- & C & M_+\\
B & M_+ & A_+
\end{pmatrix}.
\label{eq:biased_transfer_matrix}
\end{equation}
This matrix no longer decomposes exactly into symmetric and antisymmetric
sectors because the intrinsic chiral field removes the exact
$+\leftrightarrow-$ symmetry.  The modes describing helical occupancy and
handedness are therefore no longer exact symmetry eigenmodes and become
coupled.  Nevertheless, their physical roles remain distinct:
$\Delta_{\chi}$ selects the sign of the preferred handedness, while the wall
free energy
\[
\Delta F_{\rm wall}=K+J
\]
suppresses reversals of that sign.

\subsection{Finite-Domain Amplification of Handedness}

The amplification of a small intrinsic bias can be seen by comparing two
otherwise equivalent homochiral helical domains of length $L$, one
right-handed and one left-handed.  This is the finite-domain analogue of the
response of a ferromagnetic domain to a magnetic field.  For an Ising domain
of $L$ aligned spins, the field produces a free-energy difference
proportional to $2HL$.  Here the intrinsic molecular bias $\Delta_\chi$
plays the role of the field.  The free-energy difference between the two
homochiral domains is
\begin{equation}
F_L-F_R
\simeq
2L\Delta_{\chi}.
\label{eq:domain_bias_free_energy}
\end{equation}
The corresponding probability ratio is therefore
\begin{equation}
\frac{P_R}{P_L}
\simeq
\exp(2\beta\Delta_{\chi}L).
\label{eq:PR_PL_bias}
\end{equation}

Equation~\eqref{eq:PR_PL_bias} is not a new ferromagnetic result, but its
application to a finite protein helix is physically revealing.  The
domain-wall penalty maintains coherent handedness over the length $L$,
allowing a small residue-level stereochemical bias to accumulate extensively
over the domain.  The length $L$ should therefore be interpreted as the
length over which the handedness remains coherent.  In practice, it cannot
substantially exceed the observable homochiral length determined earlier by
the competition between $\xi_H$ and $\xi_\chi$.

For a modest local bias $\Delta_{\chi}=0.1\,k_{\rm B}T$ per residue, an
ordinary ten-residue protein helix gives
\begin{equation}
\left.
\frac{P_R}{P_L}
\right|_{L=10}
\simeq
e^2
\simeq
7.4,
\label{eq:PRPL_example_10}
\end{equation}
whereas a 20-residue helical domain gives
\begin{equation}
\left.
\frac{P_R}{P_L}
\right|_{L=20}
\simeq
e^4
\simeq
55.
\label{eq:PRPL_example_01}
\end{equation}
For the same local bias, a 39-residue alanine-based helix gives
\begin{equation}
\left.
\frac{P_R}{P_L}
\right|_{L=39}
\simeq
e^{7.8}
\simeq
2.4\times10^{3},
\label{eq:PRPL_example_39}
\end{equation}
while a coherently handed domain of 58 residues gives
\begin{equation}
\left.
\frac{P_R}{P_L}
\right|_{L=58}
\simeq
e^{11.6}
\simeq
1.1\times10^{5}.
\label{eq:PRPL_example_58}
\end{equation}
The lengths used here correspond to experimentally relevant scales discussed
earlier: ordinary protein $\alpha$-helices are often of order ten residues,
alanine-based helix--coil experiments have examined peptides containing
19--39 residues, and the single-$\alpha$-helical domain of myosin VI extends
over approximately 58 residues
\cite{KabschSander1983,HuangGetahunZhuKallenbach2004,BarnesMyosinVI2019}.
Thus the same small local bias that produces only moderate selection in a
short helix becomes effectively decisive in a longer coherently handed
domain.

The sensitivity to the magnitude of the local bias is equally striking.  If
$\Delta_{\chi}=0.2\,k_{\rm B}T$, a 20-residue helical domain gives
\begin{equation}
\frac{P_R}{P_L}
\simeq
e^8
\simeq
3.0\times10^{3}.
\label{eq:PRPL_example_02}
\end{equation}
For $L=39$ and $L=58$, the corresponding ratios are approximately
$6.0\times10^{6}$ and $1.2\times10^{10}$, respectively.  These values
demonstrate that the preference for right-handed $\alpha$-helices in
L-amino-acid proteins need not originate from a large local free-energy
difference.  A bias substantially smaller than $k_{\rm B}T$ per residue can
be amplified strongly when it is repeated coherently over a finite helical
domain.

\subsection{Protein, Polyalanine, and Synthetic-Polymer Examples}

Poly-L-alanine provides a useful molecular limiting case because all
nonterminal residues possess the same L-alanine stereochemistry.  Classical
conformational-energy calculations found the right-handed
$\alpha$-helical conformation of poly-L-alanine to be lower in energy than
the corresponding left-handed conformation by approximately
$0.8\,{\rm kcal\,mol^{-1}}$ per residue
\cite{RamachandranPolyLAlanine1971}.  In the present convention, this
right--left separation corresponds to
\begin{equation}
2\Delta_{\chi}^{({\rm Ala})}
\simeq
0.8\,{\rm kcal\,mol^{-1}},
\qquad
\Delta_{\chi}^{({\rm Ala})}
\simeq
0.4\,{\rm kcal\,mol^{-1}}
\simeq
0.7\,k_{\rm B}T
\label{eq:polyalanine_bias}
\end{equation}
near room temperature.  This number should be interpreted as a microscopic
conformational-energy estimate for an idealized poly-L-alanine chain, rather
than as a universal solution-phase free-energy difference for every
alanine-containing protein helix.  Solvation, conformational entropy, end
effects, and the surrounding sequence can renormalize the effective
$\Delta_\chi$ entering the coarse-grained theory.  Its significance here is
that L-alanine stereochemistry itself contains a pronounced right-handed
preference, consistent with the asymmetry of the Ramachandran landscape in
Fig.~2.

Synthetic helical polymers provide a complementary realization of the same
amplification principle.  In the polyisocyanate systems studied by Green and
coworkers, a small excess of one chiral component was found to produce a much
larger preference for one macromolecular helical sense
\cite{GreenEtAl,GreenSergeants1989}.  Statistical-mechanical analyses of
these systems using one-dimensional Ising and random-field Ising models
identified a local chiral energy scale of order a fraction of
$k_{\rm B}T$ per repeat unit, together with a helix-reversal energy of several
$k_{\rm B}T$
\cite{SelingerSelinger1996,SelingerSelinger1997}.  Representative values may
be summarized as
\begin{equation}
\Delta_{\chi}^{({\rm syn})}
\sim
0.2\,{\rm kcal\,mol^{-1}},
\qquad
\Delta F_{\rm wall}^{({\rm syn})}
\sim
4\,{\rm kcal\,mol^{-1}}.
\label{eq:synthetic_polymer_scales}
\end{equation}
The precise mapping depends on the polymer architecture and on whether the
chiral units are distributed uniformly or randomly along the chain.  In a
random copolymer, the local chiral fields need not all have the same sign, so
the uniform-domain expression in Eq.~\eqref{eq:PR_PL_bias} cannot be applied
by simply replacing $L$ with the total chain length.  Nevertheless, the
physical separation of scales is the same as in the present theory: a weak
local bias selects the favored helical sense, while a substantially larger
reversal penalty permits that selection to remain correlated over many repeat
units.

The protein and synthetic-polymer examples therefore illustrate two related
but distinct realizations of chiral amplification.  In an L-amino-acid
protein or in poly-L-alanine, the sign of the local field is fixed by residue
stereochemistry and favors the right-handed $\alpha$-helical basin shown in
Fig.~2.  In suitably designed synthetic polymers, the sign and magnitude of
the field may instead be controlled by chiral pendant groups, composition, or
solvent.  In both cases, however, the local bias alone does not determine the
persistence length.  Its effect becomes strongly amplified only when the
domain-wall free energy suppresses reversals and allows the bias to accumulate
coherently along a finite domain.

This discussion clarifies the distinct roles of the three principal
free-energy contributions.  The local helix--coil free energy determines
whether a segment becomes helical.  The wall free energy $K+J$ determines
whether a helical domain remains homochiral.  The intrinsic chiral bias
$\Delta_{\chi}$ determines which handedness is selected.  In proteins,
L-amino-acid stereochemistry fixes the sign of $\Delta_{\chi}$ and favors the
right-handed $\alpha$-helix, while the domain-wall penalty preserves and
cooperatively amplifies this choice over many residues.

\section{Conclusion}

Let us first summarize the main results of the present study. We have
developed a three-state transfer-matrix theory for fluctuating chiral
domains in helical polymers. The model is obtained by coarse-graining
the Ramachandran conformational landscape into three physically
relevant states corresponding to coil and the two possible helical
handedness states. It is motivated by the classical helix--coil
theories of Zimm--Bragg, Gibbs--DiMarzio, Lifson--Roig, and
Poland--Scheraga, but extends them by retaining right- and left-handed
helical conformations explicitly
\cite{GibbsDiMarzio1958, ZimmBragg1958, LifsonRoig1961,
PolandScheraga1970}. Each segment is assigned a variable
$s_i=0,+1,-1$, where $s_i=0$ denotes coil and $s_i=\pm1$ denote the
two helical handedness states. The helical occupancy is
$n_i=s_i^2$, whereas $s_i$ itself carries the handedness. This
separation makes it possible to distinguish the formation and
termination of a helical domain from the persistence and reversal of
its chirality.

A central ingredient of the theory is the entropy of the coil state.
The coil is not a single microscopic conformation but a large
manifold of backbone, side-chain, and solvent-coupled configurations.
This manifold is represented by a configurational multiplicity $g$,
with the associated entropy $s_c=k_B\ln g$. The multiplicity is
essential: without it, a model containing two helical states and one
coil state would predict the unphysical high-temperature limit in
which the helical fraction approaches $2/3$. Inclusion of the coil
entropy restores the proper competition between the entropically
favored coil ensemble and the energetically stabilized helical
states. This construction connects naturally with the
rotational-isomeric-state description of polymer conformations
\cite{VolkensteinRIS,FloryRIS}.

The second essential step is the separation of same-handed helical
propagation from wrong-handed contact formation. A
chirality-independent attractive interaction would stabilize both
$+|+$ and $+|-$ contacts. Such a description is inappropriate when a
handedness reversal requires partial unwinding, loss of torsional
registry, disruption of hydrogen-bond continuity, packing
frustration, or local solvent reorganization. We therefore introduced
separate projectors for same-handed propagation and for the formation
of a handedness-reversal defect. The parameter $K$ denotes the
free-energy stabilization obtained by continuing a helix with the same
handedness, whereas $J$ denotes the additional mismatch penalty
associated with a wrong-handed contact. Replacing a same-handed
contact by a wrong-handed one therefore both loses the stabilization
$K$ and incurs the penalty $J$. The full chiral-domain-wall free
energy is consequently
\begin{equation}
\Delta F_{\rm wall}=K+J.
\end{equation}

The transfer matrix separates naturally into symmetric and
antisymmetric sectors. The symmetric sector describes the competition
between coil and helical occupancy and controls the ordinary
helix--coil correlation length. The antisymmetric sector describes
handedness fluctuations and controls the chiral persistence length.
The resulting length scales are
\begin{equation}
\xi_H\sim \sigma^{-1/2},
\qquad
\xi_\chi\simeq
\frac{1}{2}\exp[\beta(K+J)]
=
\frac{1}{2}\exp(\beta\Delta F_{\rm wall}) .
\end{equation}
The first is the familiar helix--coil cooperativity length, governed
primarily by the cost of nucleating and terminating a helical segment.
The second is a distinct chiral persistence length governed by the
free-energy cost of reversing handedness within an otherwise helical
domain. Their separation is the central analytical and physical
result of the present theory.

This distinction changes how finite helical structures should be
interpreted. The length of a contiguous helix need not be identical
to the length over which its handedness persists. A relatively short
helix may possess a very large intrinsic chiral persistence length but
terminate through a helix--coil boundary before a handedness reversal
is encountered. Conversely, a long helical region may contain several
right- and left-handed domains if the wall free energy is small. The
observed homochiral length is therefore determined by the competition
between helix--coil termination and chiral-wall formation, rather than
by a single cooperativity parameter.

The theory also clarifies the origin of local homochirality. In the
unbiased model, right- and left-handed helices are degenerate, and no
global handedness is imposed. A helical nucleus may initially choose
either sign. Once a handedness has been selected locally, however, a
reversal creates a domain wall and costs $\Delta F_{\rm wall}$.
Mixed-handed domains are consequently suppressed, and a finite helical
segment tends to remain homochiral over a distance of order
$\xi_\chi$. Local homochirality therefore arises from the suppression
of domain walls rather than from an assumed infinite preference for
one helical state. This interpretation connects the present framework
with broader studies of chiral amplification, preferred handedness,
and helical-sense control in synthetic macromolecules
\cite{GreenSergeants1989,YashimaChemRev2009,YashimaAccount2008,
SchwartzPolyisocyanides2011,KumakiAFM2009}.

The numerical estimates show that the required free-energy scales are
modest. A chiral persistence length of approximately $10$--$100$
segments corresponds to a wall free energy of only about
$3$--$5\,k_BT$. Such values are chemically plausible when the loss of
torsional registry, hydrogen-bond continuity, hydrophobic or aromatic
packing, side-chain contacts, and solvent organization contribute
additively. Because $\xi_\chi$ depends exponentially on
$\Delta F_{\rm wall}$, a change of only one or two $k_BT$ can produce
a large change in the extent of a homochiral domain. The corresponding
length scales cover experimentally relevant structures ranging from
ordinary short protein $\alpha$-helices to alanine-rich peptides and
longer single-$\alpha$-helical domains
\cite{KabschSander1983,Creighton1993,
HuangGetahunZhuKallenbach2004,BarnesMyosinVI2019}.

For proteins, one must further distinguish the persistence of a
selected handedness from the selection of that handedness in the first
place. The stereochemistry of the L-amino-acid backbone produces an
intrinsic local free-energy bias $\Delta_\chi$ toward the
right-handed $\alpha$-helical basin. This bias acts as an internal
chiral field, analogous to the magnetic field in a ferromagnet. It
selects the sign of the helical state, whereas the wall free energy
$K+J$ suppresses reversals and maintains coherence over the domain.
For a homochiral domain of length $L$, the accumulated free-energy
preference is proportional to $2L\Delta_\chi$, giving the
finite-domain probability ratio
\begin{equation}
\frac{P_R}{P_L}
\simeq
\exp(2\beta\Delta_\chi L).
\end{equation}
Thus even a weak residue-level stereochemical preference can become
decisive when it is repeated coherently over a sufficiently long
domain. The local helical free energy determines whether a segment
becomes helical, the wall free energy determines whether the resulting
domain remains homochiral, and the intrinsic chiral bias determines
which handedness is selected.

The dynamical consequences are richer than a single helix lifetime.
The equilibrium transfer matrix determines compositions, correlation
lengths, and wall densities, but kinetic rates require an additional
stochastic description, like a Glauber's kinetic Ising model \cite{Glauber1963}. 
A helical domain may shrink through end
fraying and inward motion of its two helix--coil boundaries. Near the
helix--coil crossover, the domain length fluctuates approximately
diffusively, giving the characteristic scaling
$\tau_H^{\rm char}\sim \xi_H^2$. Chiral memory may instead be lost
through activated nucleation of a handedness-reversal wall followed by
the motion of that wall through the surviving helical domain. The
reversal time may therefore be limited either by the waiting time for
wall creation or by the subsequent wall-traversal time. An intact
helical domain may also migrate along the sequence through coordinated
growth at one end and fraying at the other. Helix survival, wall
nucleation, wall propagation, and whole-domain migration are therefore
distinct dynamical processes and need not share the same rate-limiting
step.

The broader significance of the separation between local bias, cooperative stabilization, and defect-mediated loss of handedness is also clarified by the three perspectives discussed in the Introduction. 

(i) Efimov's analysis of protein handedness shows that handedness is not an incidental property of isolated $\alpha$-helices, but a recurring organizing principle across secondary, supersecondary, and higher-order protein structures.\cite{Efimov2018} The present theory addresses one statistical-mechanical element of that broader problem: the persistence and reversal of handedness in a finite fluctuating helical domain. 

(ii) The recent study of kinked-$\beta$ sheets by Roy, Appadurai, and Srivastava illustrates a complementary point, namely that Ramachandran-space information, while essential for identifying locally allowed conformations, must often be supplemented by collective descriptors when one wishes to characterize defects, reversibility, and dynamical structural motifs.\cite{RoyAppaduraiSrivastava2025} In the present work, the corresponding collective descriptors are the helix--coil boundary, the chiral domain wall, and the two correlation lengths $\xi_H$ and $\xi_\chi$. 

(iii) Finally, the concept of biological degeneracy introduced by Edelman and Gally provides a useful interpretation of the effective free energies used here.\cite{EdelmanGally2001} The same coarse-grained propagation and wall parameters may arise from several structurally distinct microscopic sources, including hydrogen-bond registry, torsional frustration, side-chain packing, hydrophobic association, and solvent reorganization. Thus the parameters $K$ and $J$ should be viewed not as uniquely assigned microscopic bond energies, but as emergent cooperative free energies that collect multiple routes to helix stabilization and handedness protection.


\appendix

\section{Microscopic Determination of the Effective Parameters}

The transfer-matrix parameters used in the main text are effective free
energies rather than individual microscopic bond energies.  This Appendix
summarizes their relation to familiar polymer and protein-physics quantities.

For a rotational-isomeric-state description with local torsional states
$\alpha$ of energy $E_\alpha$, the effective coil multiplicity is
\begin{equation}
g_{\rm RIS}
=
\sum_{\alpha}
\exp(-\beta E_\alpha).
\label{eq:appendix_gRIS}
\end{equation}
For one trans state of energy zero and two equivalent gauche states of
energy $\Delta E_g$,
\begin{equation}
g_{\rm RIS}
=
1+2\exp(-\beta\Delta E_g).
\label{eq:appendix_gRIS_simple}
\end{equation}
For a coarse-grained segment containing several torsional degrees of
freedom, the corresponding multiplicities multiply.

The helix--coil nucleation parameter is related to the free-energy cost
$\gamma_{\rm hc}$ of one helix--coil boundary by
\begin{equation}
\sigma
=
\exp(-2\beta\gamma_{\rm hc}).
\label{eq:appendix_sigma}
\end{equation}
The same-handed propagation parameter $K$ collects the cooperative
stabilization obtained by maintaining hydrogen-bond registry, torsional
continuity, packing, and favorable side-chain or solvent-mediated
interactions.  The parameter $J$ is the additional mismatch penalty for
placing neighboring helical units in opposite handedness.  The corresponding
chiral-wall free energy is therefore
\begin{equation}
\Delta F_{\rm wall}=K+J.
\label{eq:appendix_wall}
\end{equation}

Finally, the intrinsic chiral field is
\begin{equation}
\Delta_\chi
=
\frac{1}{2}
\left(
f_L^{\rm helix}-f_R^{\rm helix}
\right),
\label{eq:appendix_chiral_field}
\end{equation}
and is positive for the right-handed preference of an L-amino-acid
backbone under the convention used in the main text.  Numerical examples
for proteins, poly-L-alanine, and synthetic polymers are given in
Sec.~\ref{sec:intrinsic_chiral_bias}.


\textbf{ACKNOWLEDGMENTS}

It is a pleasure to thank Professor Anand Srivastava  for discussions and for suggesting important
references for this work.


\end{document}